\documentclass[sigconf]{acmart}

\PassOptionsToPackage{protrusion, expansion}{microtype} 

\usepackage{amsmath, amsthm}
    
\usepackage{graphicx, xcolor} 
\usepackage{hyperref, url}
\usepackage{array}
\usepackage{booktabs, multirow}
\usepackage{caption}
\usepackage{subcaption}
\usepackage{enumitem}
\usepackage{setspace}

\usepackage[ruled, linesnumbered, norelsize]{algorithm2e}

\setcopyright{rightsretained}

\copyrightyear{2023}
\acmYear{2023}
\setcopyright{licensedusgovmixed}\acmConference[SC-W 2023]{Workshops of
The International Conference on High Performance Computing, Network,
Storage, and Analysis}{November 12--17, 2023}{Denver, CO, USA}
\acmBooktitle{Workshops of The International Conference on High Performance
Computing, Network, Storage, and Analysis (SC-W 2023), November 12--17,
2023, Denver, CO, USA}
\acmPrice{15.00}
\acmDOI{10.1145/3624062.3625123}
\acmISBN{979-8-4007-0785-8/23/11}

\begin{document}

\title{Analyzing Impact of Data Reduction Techniques on Visualization for AMR Applications Using AMReX Framework}

\settopmatter{authorsperrow=3}

\settopmatter{authorsperrow=3}

\newcommand{\iu}{Indiana University}
\newcommand{\anl}{Argonne National Laboratory}
\newcommand{\lanl}{Los Alamos National Laboratory}
\newcommand{\lbl}{Lawrence Berkeley National Laboratory}
\newcommand{\pnnl}{Pacific Northwest National Laboratory}
\newcommand{\fsu}{Florida State University}

\newcommand{\AFFIL}[4]{%
     \affiliation{%
         \institution{#1}
         \city{#2}\state{#3}\country{#4}
     }
     }

\newcommand{\IU}{\AFFIL{\iu}{Bloomington}{IN}{USA}}
\newcommand{\ANL}{\AFFIL{\anl}{Lemont}{IL}{USA}}
\newcommand{\LANL}{\AFFIL{\lanl}{Los Alamos}{NM}{USA}}
\newcommand{\LBL}{\AFFIL{\lbl}{Berkeley}{CA}{USA}}
\newcommand{\PNNL}{\AFFIL{\pnnl}{Richland}{WA}{USA}}
\newcommand{\FSU}{\AFFIL{\fsu}{Tallahassee}{FL}{USA}}

\author{Daoce Wang}{\IU}
\email{daocwang@iu.edu}

\author{Jesus Pulido}{\LANL}
\email{pulido@lanl.gov}

\author{Pascal Grosset}{\LANL}
\email{pascalgrosset@lanl.gov}

\author{Jiannan Tian}{\IU}
\email{jti1@iu.edu}

\author{James Ahrens}{\LANL}
\email{ahrens@lanl.gov}

\author{Dingwen Tao}{\IU}
\email{ditao@iu.edu}
\begin{abstract}

Today’s scientific simulations generate exceptionally large volumes of data, challenging the capacities of available I/O bandwidth and storage space. This necessitates a substantial reduction in data volume, for which error-bounded lossy compression has emerged as a highly effective strategy. A crucial metric for assessing the efficacy of lossy compression is visualization. Despite extensive research on the impact of compression on visualization, there is a notable gap in the literature concerning the effects of compression on the visualization of Adaptive Mesh Refinement (AMR) data. AMR has proven to be a potent solution for addressing the rising computational intensity and the explosive growth in data volume that requires storage and transmission. However, the hierarchical and multi-resolution characteristics of AMR data introduce unique challenges to its visualization, and these challenges are further compounded when data compression comes into play. This article delves into the intricacies of how data compression influences and introduces novel challenges to the visualization of AMR data.

\end{abstract}

\maketitle

\pagestyle{plain}
\setlength{\textfloatsep}{6pt}
\section{Introduction}
\label{sec:introduction}

In recent years, there has been a significant surge in both the scale and expense associated with scientific simulations. To mitigate this, numerous high-performance computing (HPC) simulation packages, such as AMReX \cite{zhang2019amrex} and Athena++ \cite{stone2020athena++}, have incorporated Adaptive Mesh Refinement (AMR) as a strategy to reduce computational demands while preserving or even enhancing the accuracy of simulation outcomes. Unlike conventional uniform mesh methods, which maintain a consistent resolution across the entire simulation domain, AMR provides a more resource-efficient alternative by dynamically modifying the resolution, concentrating higher resolution in key areas, and thereby optimizing computational resources and storage requirements.

While AMR data can diminish the size of the output data, the reduction may not be significant enough for scientific simulations, leading to elevated I/O and storage costs. For instance, an AMR simulation with a resolution of $4096^3$ (i.e., $0.5 \times 2048^3$ mesh points at the coarse level and $0.5 \times 4096^3$ at the fine level) could produce up to 8 TB of data for a single snapshot with all data fields dumped; this translates to a necessity for 1 PB of disk storage, assuming the simulation is conducted in an ensemble of five times with 25 snapshots dumped per simulation, a common practice.

To address this, data compression methods can be employed alongside AMR techniques to further minimize I/O and storage expenses. However, traditional lossless compression methods are only marginally effective in compressing the vast volumes of data generated by scientific simulations, typically achieving a maximum compression ratio of 2$\times$.
As an alternative, a new wave of error-bounded lossy compression techniques, such as SZ~\cite{tao2017significantly, di2016fast, sz18}, ZFP~\cite{zfp}, MGARD \cite{ainsworth2018multilevel}, and their GPU versions \cite{tian2020cusz,tian2021optimizing, cuZFP}, have gained popularity in the scientific community ~\cite{di2016fast,tao2017significantly,zfp,sz18,lu2018understanding,luo2019identifying,tao2019optimizing,cappello2019use,jin2020understanding,grosset2020foresight,jin2022accelerating, baker2019evaluating}

While lossy compression has the ability to significantly decrease the I/O and storage costs linked with AMR simulations, research on its application in AMR simulations is sparse. Recently, two studies were conducted with the aim of developing efficient lossy compression techniques for AMR datasets. Luo et al. \cite{zMesh} proposed zMesh, a technique that rearranges AMR data across varying refinement levels into a 1D array to leverage data redundancy. However, compressing data into a 1D array with zMesh restricts the use of higher-dimension compression, leading to a loss of spatial information and data locality in higher-dimension data. On the other hand, Wang et al. \cite{wang2022tac} developed TAC, a technique intended to improve zMesh's compression quality through adaptive 3D compression. Although zMesh and TAC offer offline compression solutions for AMR data, they did not delve into in-situ compression, which could notably reduce the I/O cost. To bridge this gap, Wang et al. further introduced AMRIC~\cite{amric}, an in situ AMR compression framework, subsequently minimizing both I/O costs and enhancing compression quality for AMR applications.

Despite the existence of various AMR data compression solutions, none of the studies have comprehensively examined the impact of lossy compression on the visualization of AMR data. While there are studies that analyze the effects of data compression on non-AMR data visualization~\cite{2022-vis}, the visualization of AMR data is more complex due to its hierarchical structure. 

\label{sec:back-crack}
\begin{figure*}[h]
     \centering
     \begin{subfigure}[t]{0.33\linewidth}
         \centering
         \includegraphics[width=\linewidth]{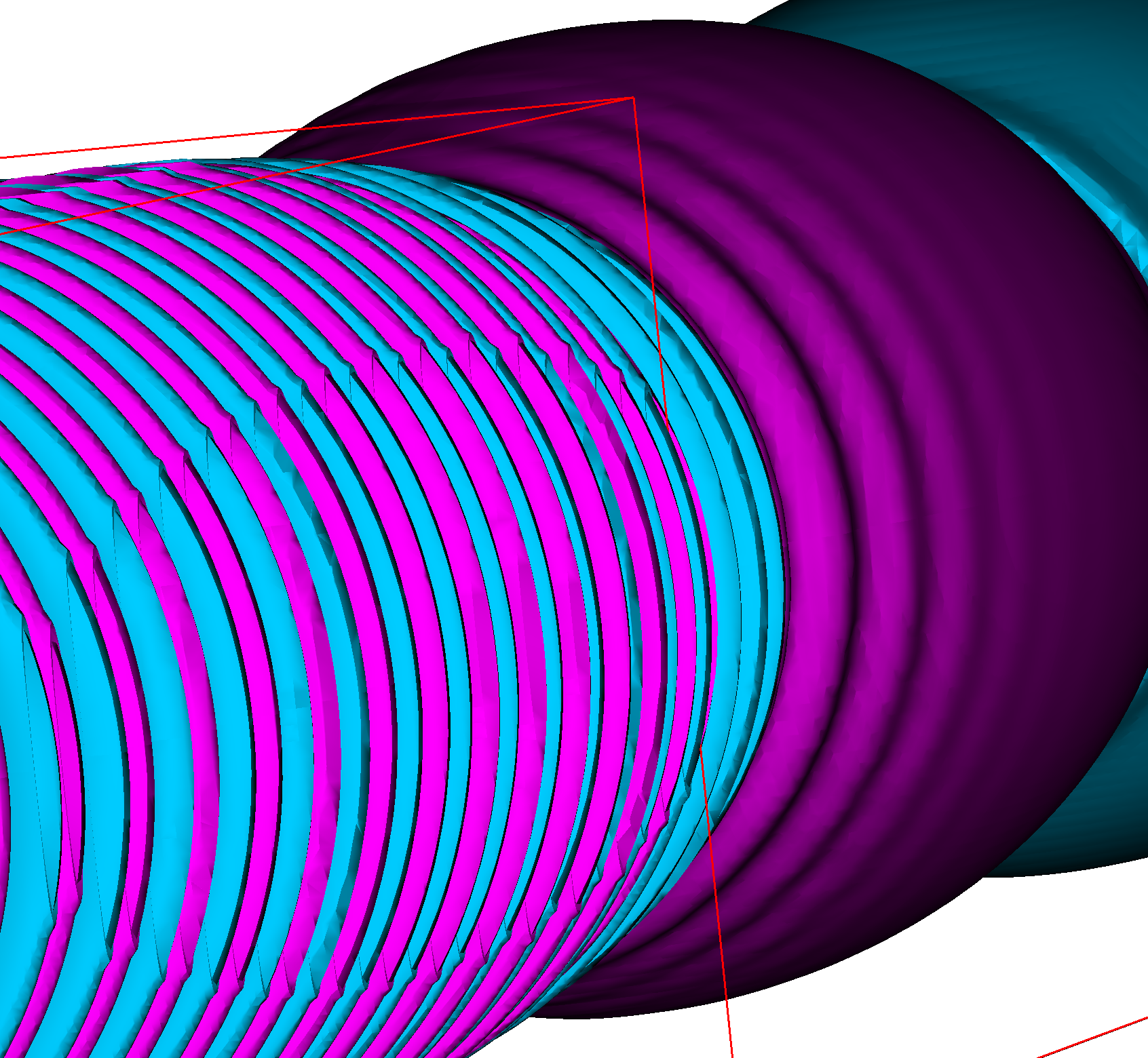}  
         \caption[t]{Original re-sampling}
         \label{fig:defualt-ori}
     \end{subfigure} 
     \begin{subfigure}[t]{0.33\linewidth}
         \centering
         \includegraphics[width=\linewidth]{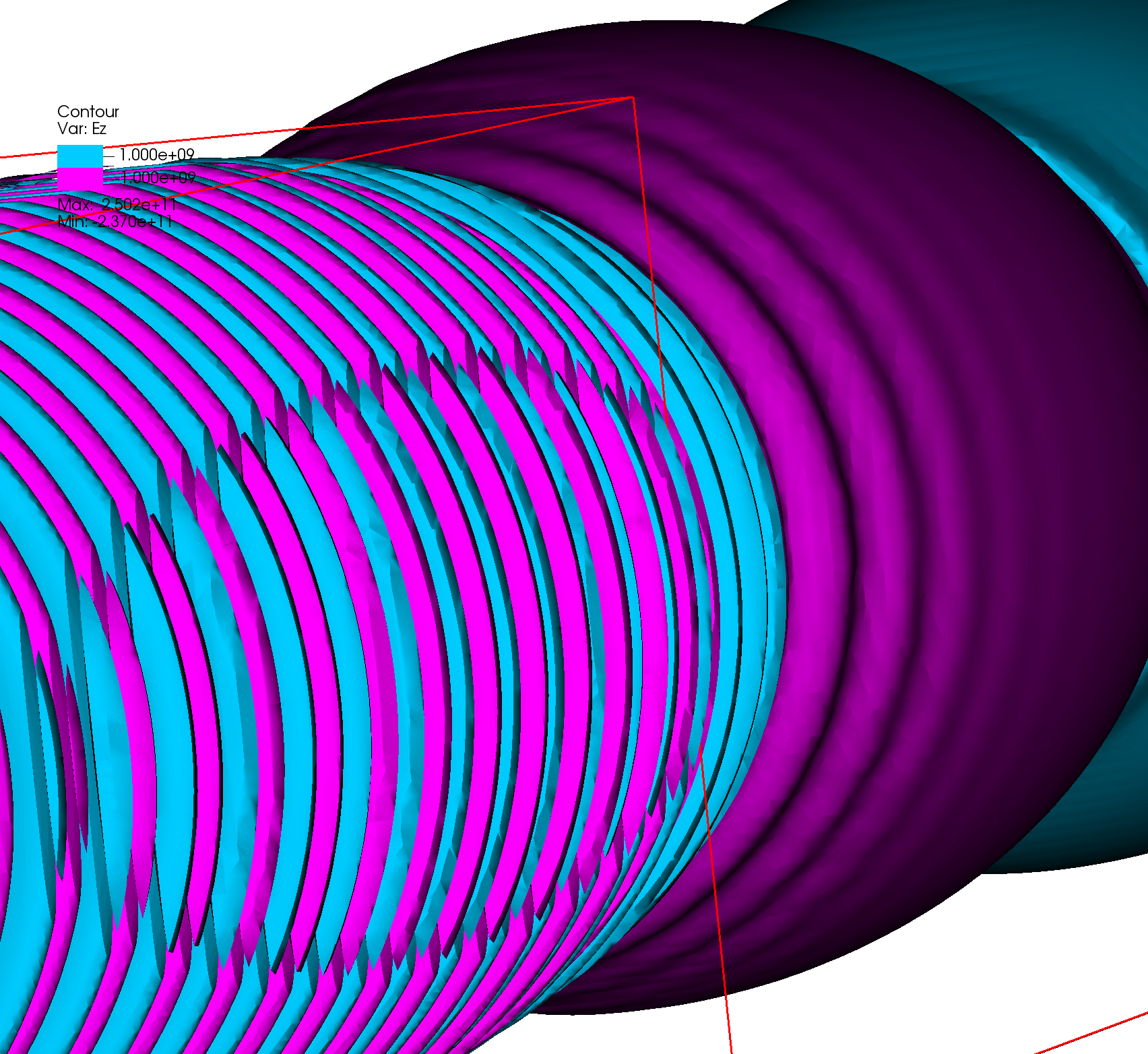}
         \caption{Dual-cell}
         \label{fig:dual-gap-ori}
     \end{subfigure} 
     \begin{subfigure}[t]{0.33\linewidth}
         \centering
         \includegraphics[width=\linewidth]{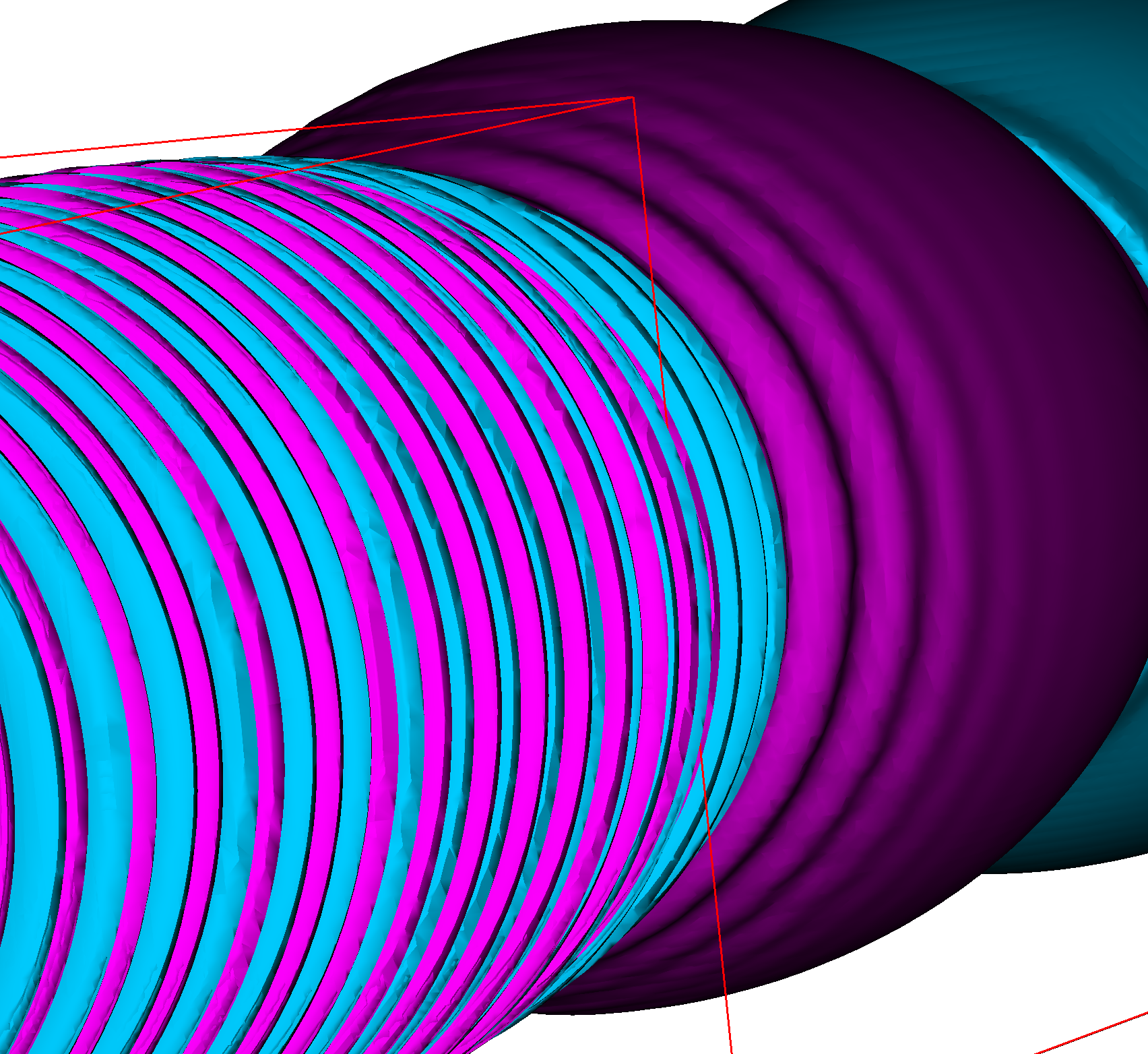}
         \caption{Dual-cell with switching cells}
         \label{fig:dual-ori}
     \end{subfigure}
     \vspace{-2mm}
        \caption[t]{Original AMR data (without compression) iso-surface visualization using marching cubes algorithm using re-sampling, dual-cell, and dual-cell with switching cells (or redundant coarse data). The red box indicates the finer level.}
        \label{fig:all}
\end{figure*}
Specifically, the original (non-compressed) AMR data presents distinctive challenges when using basic visualization methods, particularly at the intersections of different grids with varying resolutions. For instance, as illustrated in Figure~\ref{fig:defualt-ori}, the traditional re-sampling method leads to cracks between different AMR levels. However, there are advanced AMR visualization solutions designed to address this issue. For example, utilizing a dual-cell approach with switching cells can mend the cracks between different AMR levels, as depicted in Figure~\ref{fig:dual-ori}. A comprehensive explanation of the re-sampling and dual-cell methods will be provided in Sections~\ref{sec:re-mc} and~\ref{sec:dualcell}, respectively.

To gain a deeper insight into how compression influences AMR data visualization methods, this study:

\begin{itemize}
\item Apply data compression on both basic (re-sampling) and advanced (dual-cell) AMR data visualization methods,  illustrating that advanced AMR visualization methods are incompatible with, and negatively impacted by, data compression compared to basic AMR visualization methods
\item Analyze and provide a preliminary rationale for why advanced AMR visualization methods diminish the quality of decompressed data visualization relative to basic AMR visualization methods.
\end{itemize}
\section{Background And Motivation}
\label{sec:background}

In this section, we first provide background information on lossy compression techniques as they pertain to scientific data, followed by an introduction to AMR methods and the structure of AMR data. Subsequently, we discuss a common challenge faced when visualizing original, uncompressed AMR data and delve into existing solutions that have been developed to address this issue.

\subsection{Lossy Compression for Scientific Data}
\label{sec:backlc}

Lossy compression is a widely used data reduction technique that attains high compression ratios by forfeiting some non-essential information in the reconstructed data. This method is often preferred over lossless compression, particularly for continuous floating-point data, as it typically yields significantly higher compression ratios. The effectiveness of lossy compression is generally assessed using three primary metrics: compression ratio, data distortion, and compression throughput. The compression ratio is the ratio between the original and compressed data sizes, data distortion evaluates the quality of the reconstructed data in comparison to the original data using metrics like structural similarity index measure (SSIM), and compression throughput denotes the volume of data that can be compressed by the compressor within a specified time frame.

In recent times, there has been a surge in the development of high-accuracy lossy compressors specifically designed for scientific floating-point data, such as SZ~\cite{di2016fast, tao2017significantly, sz18} and ZFP~\cite{zfp}. SZ is a prediction-based compressor, while ZFP is a transform-based compressor. Both are tailored to compress scientific floating-point data and offer a precise error-controlling mechanism based on user specifications. For instance, in the error-bounded mode, users must specify a type of error-bound, like absolute error bound, and a bound value. The compressor then ensures that the discrepancies between the original and reconstructed data do not surpass the specified error bound.

In this study, we used the SZ lossy compressor framework due to its superior compression ratio and modular structure. Moreover, the SZ framework comprises various algorithms to meet diverse user requirements. For instance, SZ with Lorenzo predictor \cite{sz17} delivers high compression throughput, whereas SZ with spline interpolation \cite{sz3} yields a high compression ratio, especially for large error bounds. Generally, there are three crucial stages in prediction-based lossy compression, such as SZ. Initially, each data point's value is predicted based on its neighboring points using an optimal prediction method. Next, the difference between the actual and predicted values is quantized based on the user-defined error bound. Lastly, customized Huffman coding and lossless compression are employed to achieve high compression ratios.

Previous studies have examined the impact of lossy compression on reconstructed data quality and subsequent analysis, offering recommendations for setting compression configurations for specific applications~\cite{jin2020understanding,jin2021adaptive,sz3,sz18,sz17,sz16,jin2022accelerating}.
Additionally, a past study investigated the influence of lossy compression on the volume rendering of non-AMR cosmology data~\cite{2022-vis}.

However, there has been no comprehensive study on the impact of lossy compression on AMR data visualization. Hence, this paper undertakes a thorough investigation of how, and explains why, various types of lossy compression affect different AMR data visualization methods

\subsection{AMR Methods and AMR Data}
\label{sec:backamr}

\begin{figure}[t]
     \centering
      \includegraphics[width=0.80\linewidth]{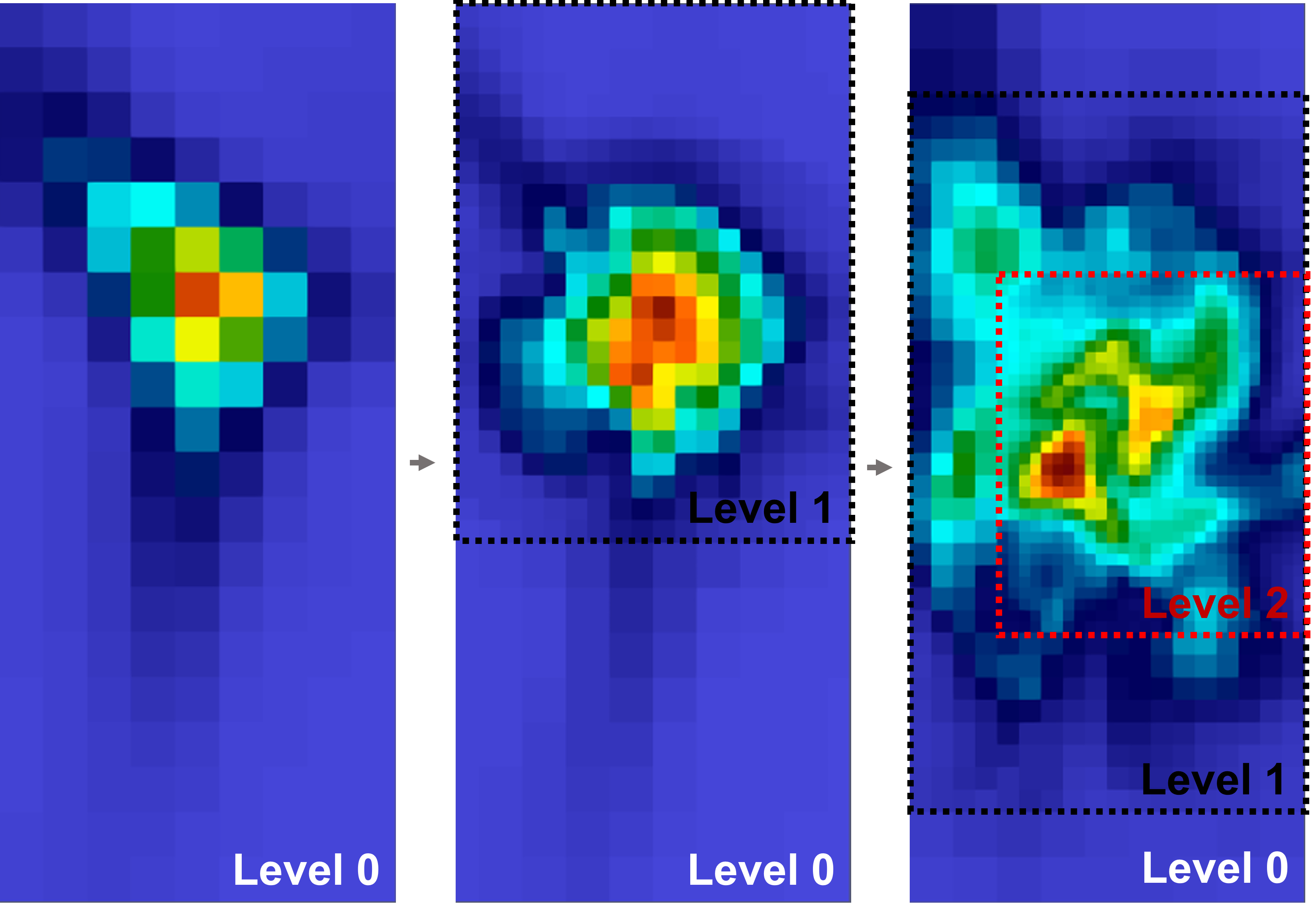}
        \caption[t]{Visualization of a zoom-in 2D slice from three pivotal timesteps created by an AMR-based cosmology simulation, Nyx. As the universe evolves, the grid structure adjusts accordingly. The dashed black and red boxes highlight areas of finer and the finest refinement, respectively.}
        \label{fig:visamr}
\end{figure}

AMR is a method that customizes the precision of a solution by utilizing a non-uniform grid, thereby enabling computational and storage efficiency without sacrificing the required accuracy. In AMR applications, the mesh or spatial resolution is modified according to the refinement level needed for the simulation. This entails employing a finer mesh in areas of higher importance or interest and a coarser mesh in less significant regions. During an AMR simulation, meshes are refined based on specific refinement criteria, such as refining a block when its norm of the gradients or maximum value surpasses a threshold, as depicted in Figure~\ref{fig:visamr}.
By dynamically adjusting the mesh resolution to meet the simulation's demands, AMR efficiently balances computational efficiency and solution accuracy, making it a potent approach for various scientific simulations.

\begin{figure}[t]
    \centering 
    \includegraphics[width=0.99\columnwidth]{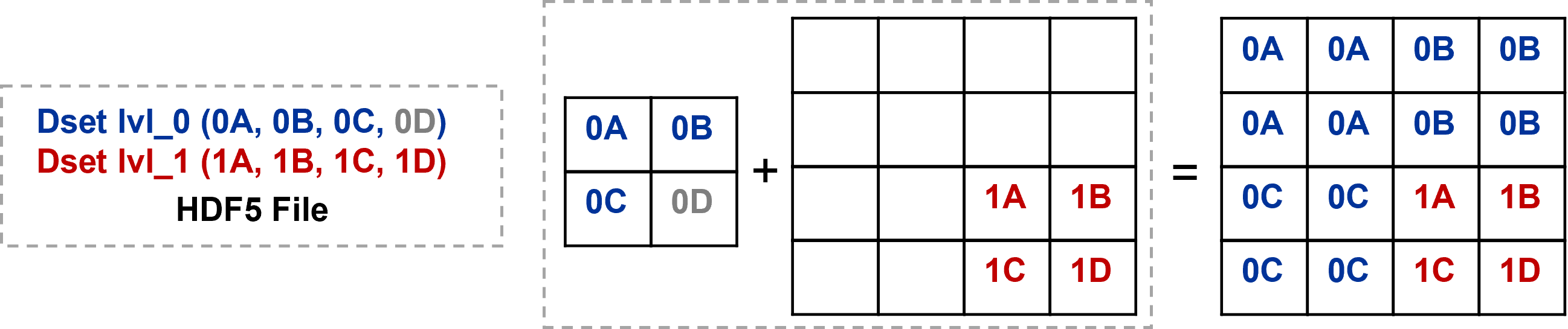}
    \caption{A typical example of AMR data storage and usage.}
    \label{fig:base_ex}
\end{figure}

AMR application-generated data is intrinsically hierarchical, with each AMR level having different resolutions. Generally, the data for each level is stored separately, for instance, in distinct HDF5 datasets (groups). For example, Figure~\ref{fig:base_ex} (left) shows a basic two-level patch-based AMR dataset in the file structure, where "0" denotes the coarse level (low resolution), and "1" represents the fine level (high resolution). When users require AMR data for post-analysis, they typically transform the data from different levels into a uniform resolution. In the provided example, the coarse-level data would be up-sampled and merged with the fine-level data, omitting the redundant coarse data point "0D," as shown in Figure~\ref{fig:base_ex} (right). This approach can also be used to visualize AMR data without requiring specific AMR visualization tool kits.

Two primary methods exist for representing AMR data: patch-based AMR and tree-based AMR~\cite{wang2020cpu}. The fundamental difference between these techniques lies in their treatment of data redundancy across various refinement levels. Patch-based AMR retains redundant data at the coarse level, as it stores data blocks to be refined at the next level within the current level, simplifying the computation involved in the refinement process. On the other hand, tree-based AMR arranges grids on tree leaves, eliminating redundant data across levels. However, tree-based AMR data can be more challenging for post-analysis and visualization compared to patch-based AMR data~\cite{harel2017two}. In this study, we concentrate on the state-of-the-art patch-based AMR framework, AMReX.


It's important to highlight that in AMReX (patch-based AMR), the redundant data from the coarser levels is frequently not used during post-analysis and visualization, as illustrated in Figure~\ref{fig:base_ex} (the coarse point "0D" is not utilized). Hence, one can omit this redundant data during the compression process to enhance the compression ratio.

\subsection{Isosurface, Re-sampling and Marching cubes}
\label{sec:re-mc}
\begin{figure}[t]
    \centering 
    \vspace{-6mm}
    \includegraphics[width=0.8\columnwidth]{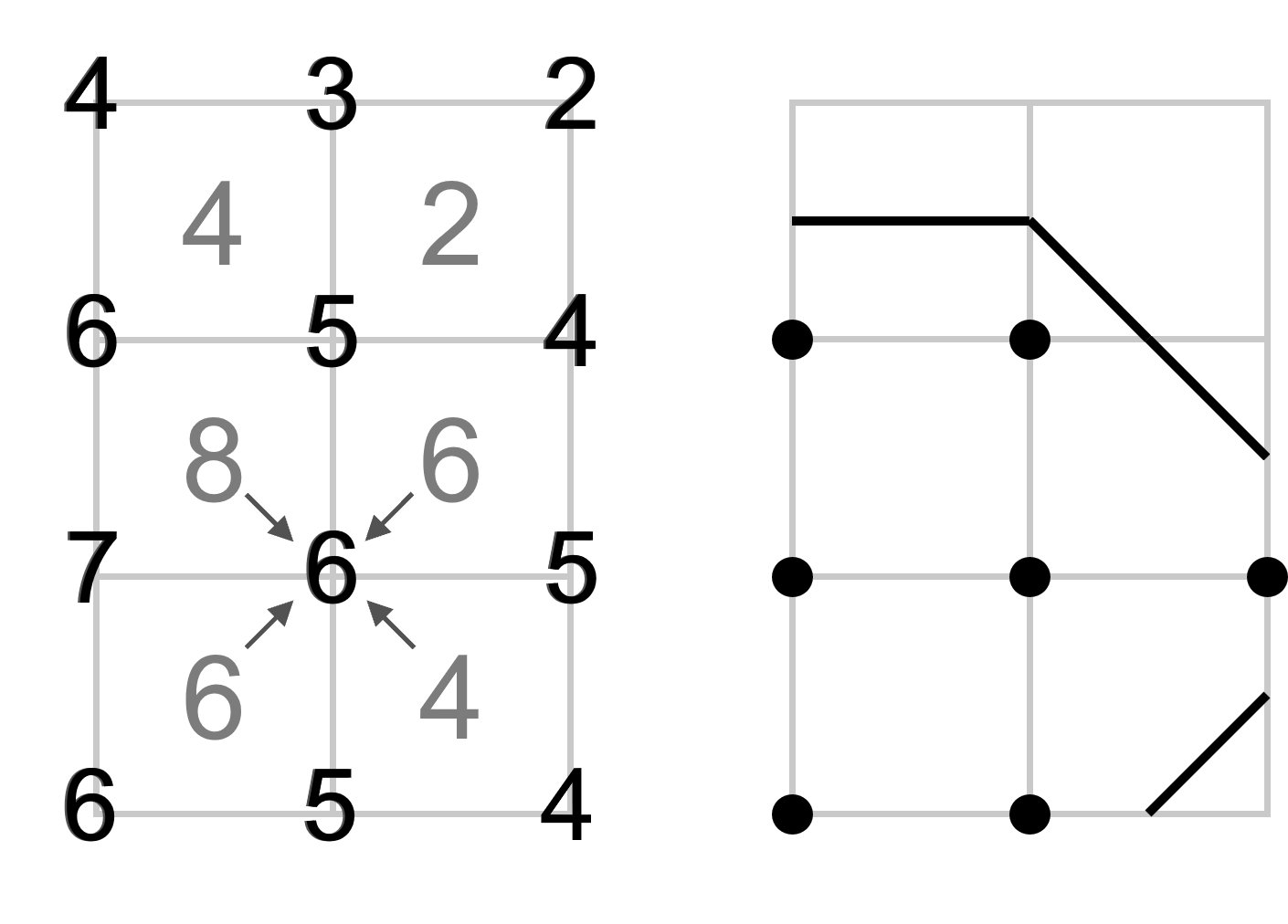}
    \caption{A 2D example of re-sample cell data to Vertex data (left) and the Marching cubes algorithm on the vertex-center data (right).}
    \vspace{-2mm}
    \label{fig:re-mc}
\end{figure} 
An isosurface represents a three-dimensional surface consisting of points that share a constant value, such as pressure, temperature, velocity, or density, within a spatial volume. This visualization technique is commonly employed in scientific research, medical imaging, and other fields requiring the analysis of volumetric data.
In iso-surface visualization, it's standard to first (1) transition from cell-centered to vertex-centered data via re-sampling and then (2) implement the Marching Cubes algorithm.

The re-sampling essentially "diffuses" the data value from the cell to its vertex using either bi-linear or tri-linear interpolation, contingent on the data's dimensionality. Figure~\ref{fig:re-mc} (left) illustrates that during this process, a vertex's value is derived from its adjacent cell's value. For example, the 6 in the middle column was calculated using its neighbor cell (i.e., 8, 6, 6, and 4).
Post-interpolation, each dimension of the vertex data size will increase by one relative to the original cell-centered data.

The Marching Cubes algorithm~\cite{m-cubes} is a foundational method for extracting iso-surfaces or iso-contours from 3D or 2D datasets. It sequentially constructs the iso-surface or contour, examining each data cube in light of varying node configurations. Given a designated iso value, a node might either surpass or not achieve this threshold. This duality results in two possible configurations per node. Consequently, a 3D cube with 8 nodes has 256 unique configurations (i.e. $2^8$), while a 2D square with 4 nodes offers 16 (i.e. $2^4$). Figure~\ref{fig:re-mc} (right)  provides a 2D example of an iso-contour whose value is equal to five.

\begin{figure}[h]
     \centering
     \begin{subfigure}[t]{0.35\linewidth}
         \centering
         \includegraphics[width=\linewidth]{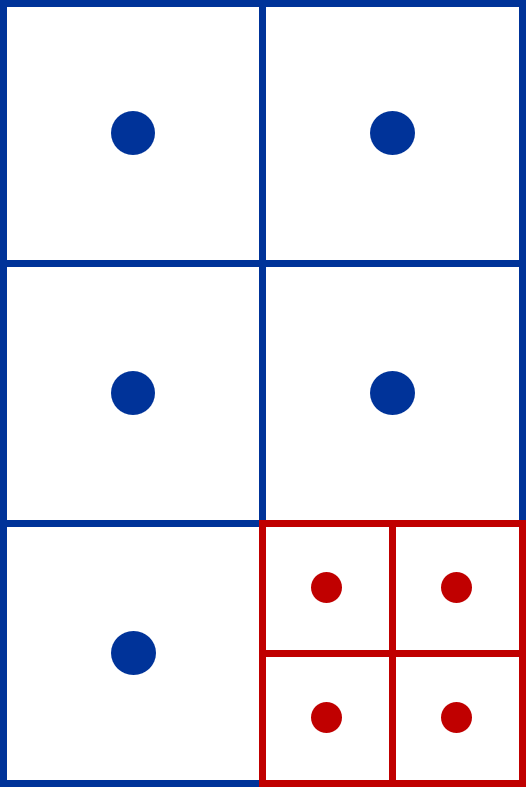}  
         \caption[t]{Cell-center AMR data}
         \label{fig:cell-amr}
     \end{subfigure} 
     \hspace{0.1\linewidth}
     \begin{subfigure}[t]{0.35\linewidth}
         \centering
         \includegraphics[width=\linewidth]{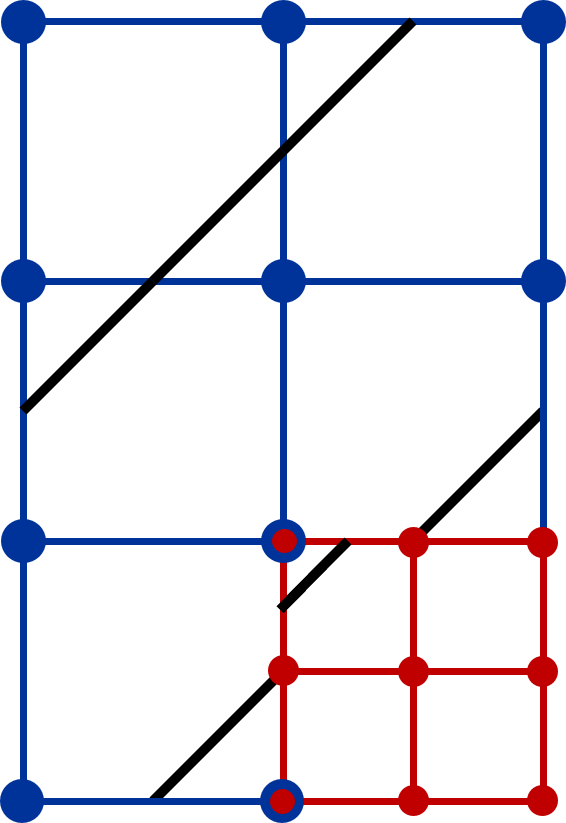}
         \caption{Vertex-center AMR data}
         \label{fig:vertex-amr}
     \end{subfigure} 
        \caption[t]{2D example of cell center AMR data and vertex-center AMR data, the black line indicates the isosurface/contour. There will be discontinuities in the isosurface between different levels due to the resolution change.}
        \label{fig:cell-vertex-amr}
\end{figure}

\begin{figure}[h]
    \centering 
    \includegraphics[width=0.8\columnwidth]{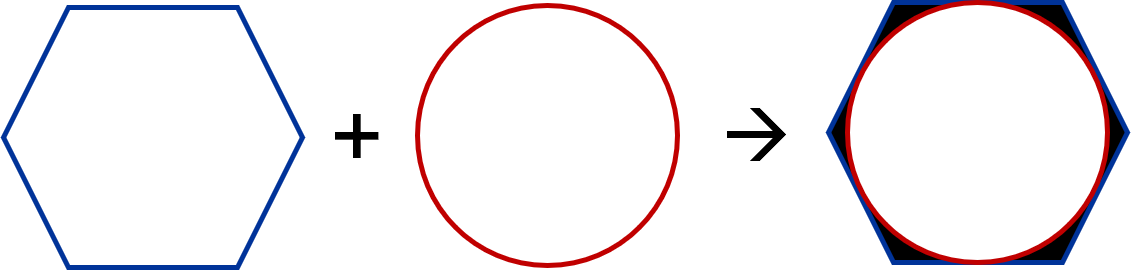}
    \caption{An intuitive 2D example for the crack/hole between different AMR levels, the blue hexagon representing the top-down view of the iso-surface of coarser, low-resolution level, and the red circle indicates the top-down view of the iso-surface of fine level with higher resolution. There will be cracks/holes when the two levels are combined.}
    \label{fig:small-example}
\end{figure} 

\subsection{Basic AMR visualization and challenge}
\label{sec:dualcell}
One common challenge in AMR data visualization is the appearance of artifacts/cracks between various AMR levels when visualizing iso-surfaces, as depicted in Figure~\ref{fig:defualt-ori}. These discrepancies arise because re-sampled vertex-centered data lead to conflicts between AMR levels, specifically, dangling nodes. This conflict results in discontinuities or cracks in the iso-surfaces or contours generated by the marching cube algorithms, as demonstrated in Figure~\ref{fig:cell-vertex-amr} and~\ref{fig:small-example} as examples.

\begin{figure}[h]
    \centering 
    \includegraphics[width=0.7\columnwidth]{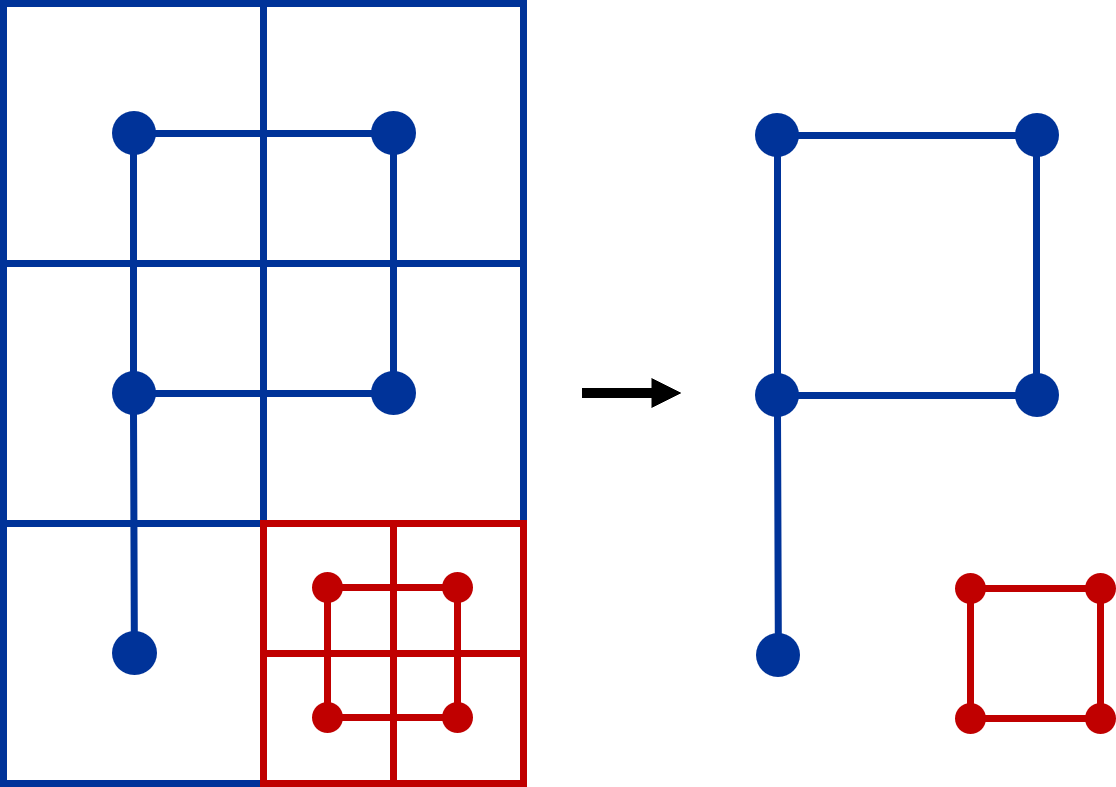}
    \caption{2D example of dual cell method}
    \label{fig:dual-cell-ex}
\end{figure}

\begin{figure}[t]
    \centering 
    \includegraphics[width=0.85\columnwidth]{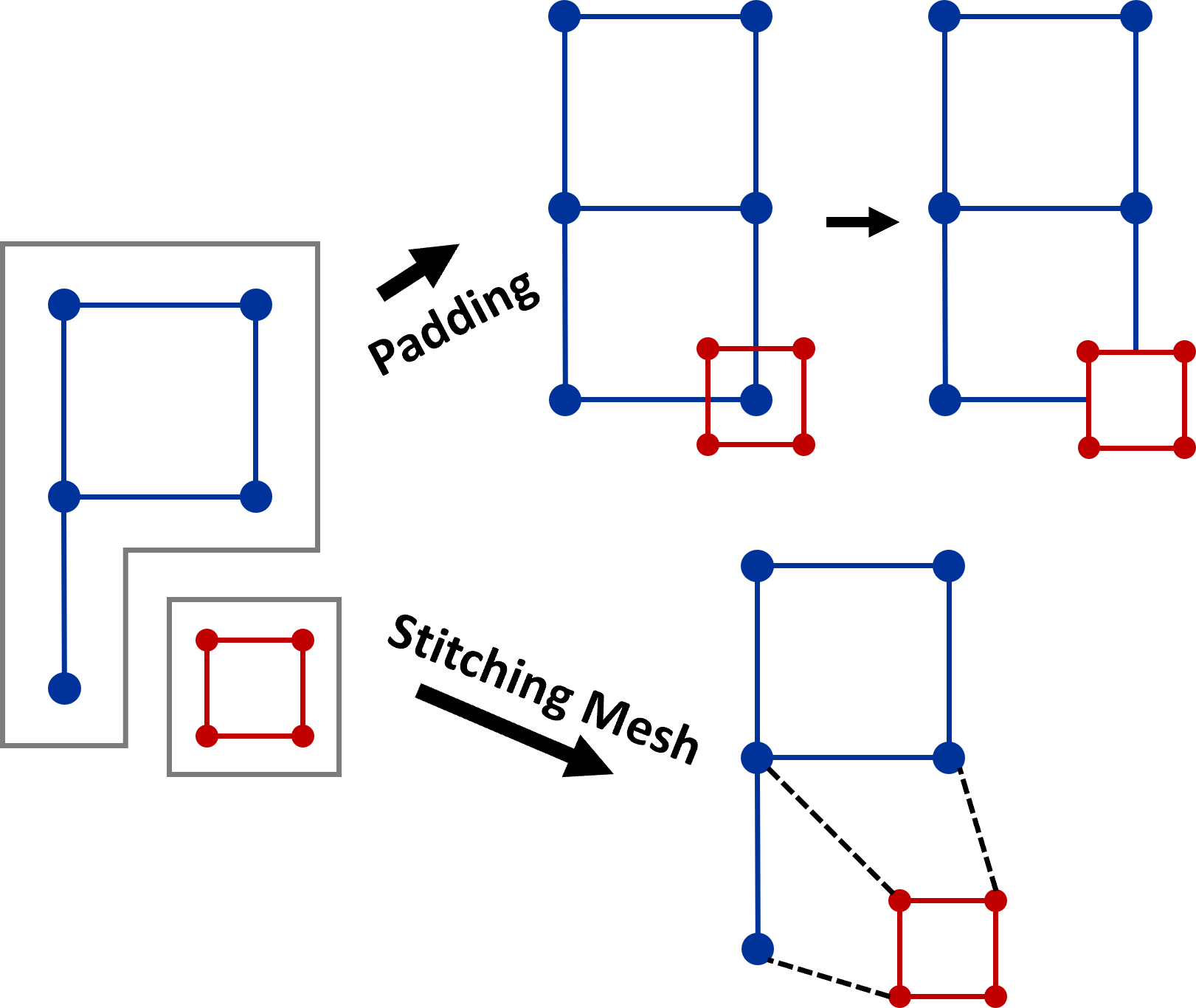}
    \caption{2D example of the gap between different levels (left) and using padding and stitching mesh to fill the gap.}
    \label{fig:gap}
\end{figure}

A standard approach to bridging these cracks across AMR levels is the adoption of dual-cell geometry~\cite{gunther-amr}. Instead of re-sampling, this method constructs a grid by connecting cell centers, as illustrated in the 2D example in Figure~\ref{fig:dual-cell-ex}. Using the original data values, the dual method mitigates the issues of conflicts between AMR levels, like dangling nodes. However, it does introduce a new challenge: gaps between hierarchy levels, as evidenced in Figure~\ref{fig:dual-gap-ori}. A 2D example of the gap is further shown on the left side of Figure~\ref{fig:gap}.

Addressing this, one might employ the stitching cells algorithm mentioned in~\cite{gunther-amr}, which utilizes stitching cells to construct an unstructured geometry across AMR levels, effectively filling the gap. This is demonstrated in the 2D sample in the lower part of Figure~\ref{fig:gap}. An alternative solution capitalizes on the existing overlap between AMR levels, especially in patch-based AMR as discussed in Section~\ref{sec:backamr}. The redundant coarse data, such as the 0D data point illustrated in Figure~\ref{fig:base_ex}, can be used to bridge the gap as shown in the upper part of Figure~\ref{fig:gap}. With either the stitching cells approach or the incorporation of redundant coarse data, gaps between AMR levels are eliminated, as seen in Figure~\ref{fig:dual-ori}. Importantly, both methods will fix the cracks that appear in Figure~\ref{fig:defualt-ori} with re-sampling.


\section{Methodology}
\subsection{Experiment overview}
In this study, we concentrate on the iso-surface visualization of compressed AMR data, as it poses greater challenges compared to other visualization techniques, such as volume rendering and slicing. Additionally, iso-surfaces are highly sensitive to errors and can be significantly affected by compression errors. This sensitivity serves as a useful lens through which we can gain a deeper understanding of the impact of compression on the data.

We conducted the visualization of the AMR iso-surface on data compressed using two distinct compression algorithms. Additionally, we utilized two different AMR applications, details of which will be elaborated in the subsequent paragraphs. Furthermore, we employed two different AMR iso-surface visualization techniques: the conventional resampling method coupled with marching cubes, and the advanced dual-cell method, also paired with marching cubes. This latter method is capable of fixing the cracks between different levels.

Except for the visualization quality, we will report quantitative metrics including the compression ratio, structural similarity index measure (SSIM), and peak signal-to-noise ratio (PSNR).

\subsection{AMR applications}
\begin{table}[h]
\caption{Detailed information about our tested AMR dataset.}
\vspace{-2mm}
\resizebox{.99\linewidth}{!}{
\begin{tabular}{|c|c|c|c|}
\hline
Runs &
  \#AMR Levels &
  \begin{tabular}[c]{@{}c@{}}Grid size of each level \\ (coarse to fine)\end{tabular} &
  \begin{tabular}[c]{@{}c@{}}Density of each level \\ (coarse to fine)\end{tabular} \\ \hline
WarpX &
  2 &
  128×128×1024, 256×256×2048 &
   91.4\%, 8.6\% \\ \hline
Nyx &
  2 &
  256×256×256, 512×512×512 &
 59.3\%,  40.7\% \\ \hline
\end{tabular}
}
\label{tab:dataset}
\vspace{-1mm}
\end{table}

\begin{figure*}[t]
     \centering
     \begin{subfigure}[t]{0.32\linewidth}
         \centering
         \includegraphics[width=\linewidth]{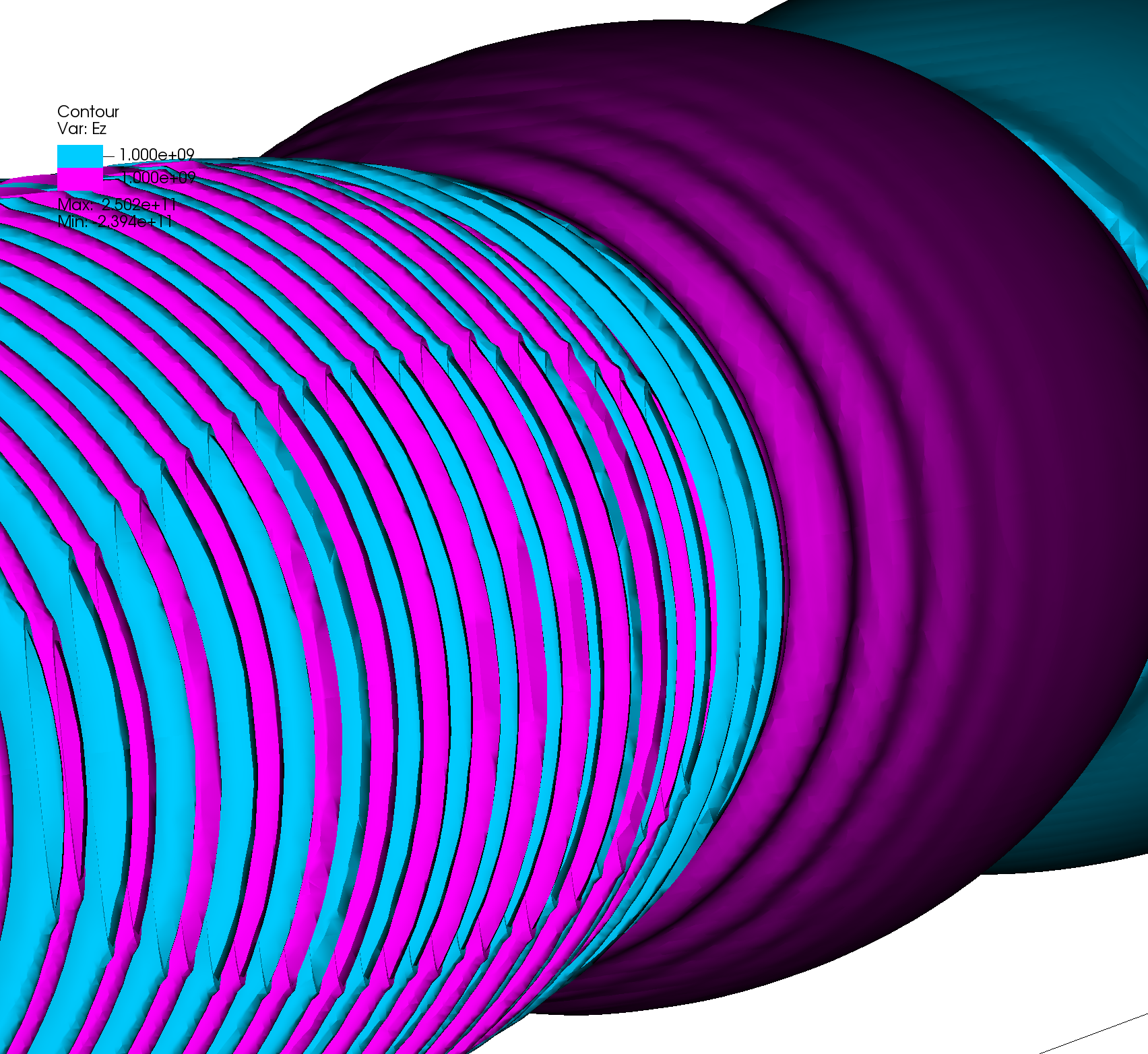}  
         \caption[t]{Re-sampling, eb=1E-4, R-SSIM=2.20E-07}
         \label{fig:wpx-dere-4}
     \end{subfigure} 
     \hspace{0.01\linewidth}
     \begin{subfigure}[t]{0.32\linewidth}
         \centering
         \includegraphics[width=\linewidth]{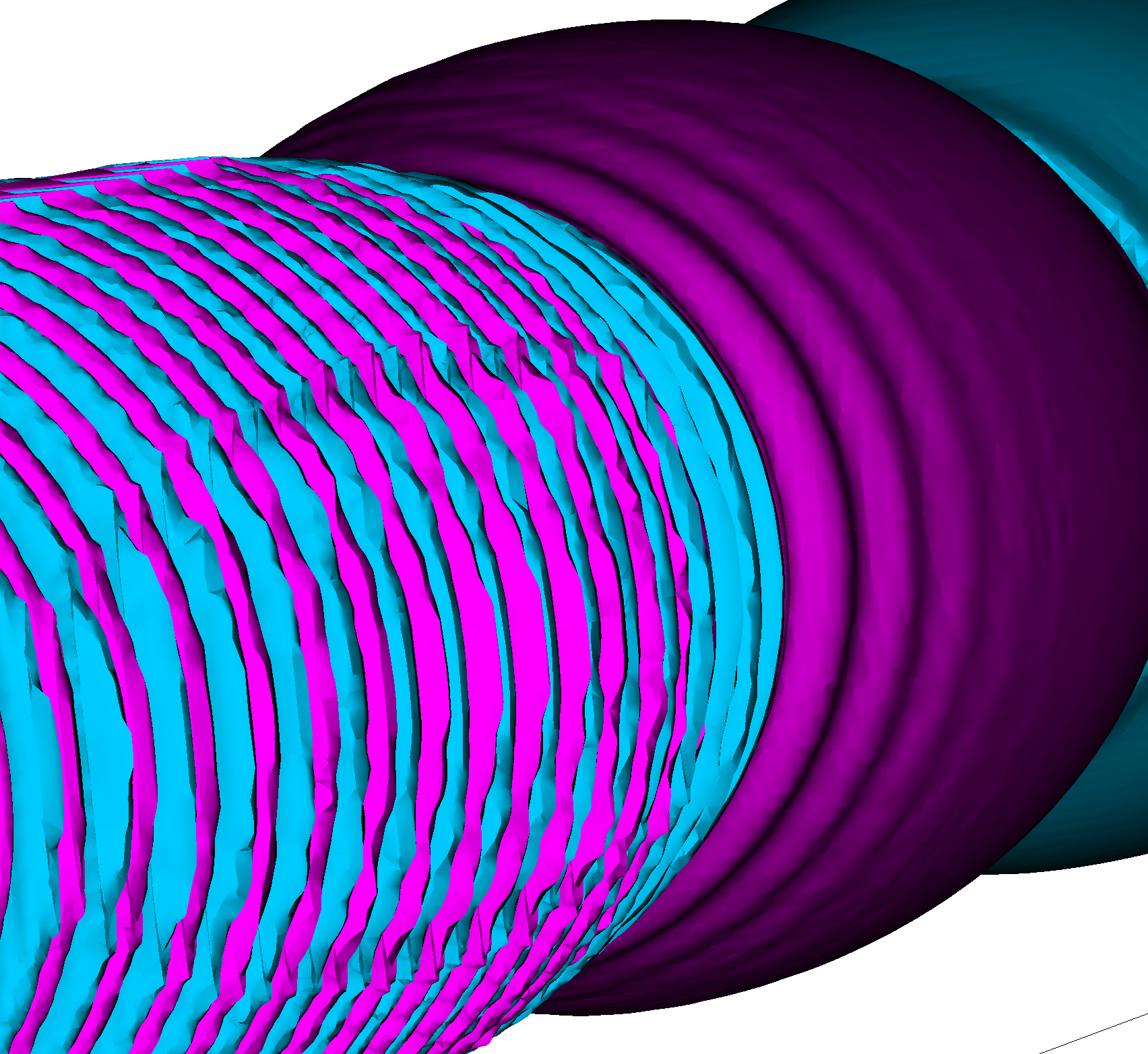}
         \caption{Re-sampling, eb=1E-3, R-SSIM=1.38E-05}
         \label{fig:wpx-dere-3}
     \end{subfigure} 
     \hspace{0.01\linewidth}
     \begin{subfigure}[t]{0.32\linewidth}
         \centering
         \includegraphics[width=\linewidth]{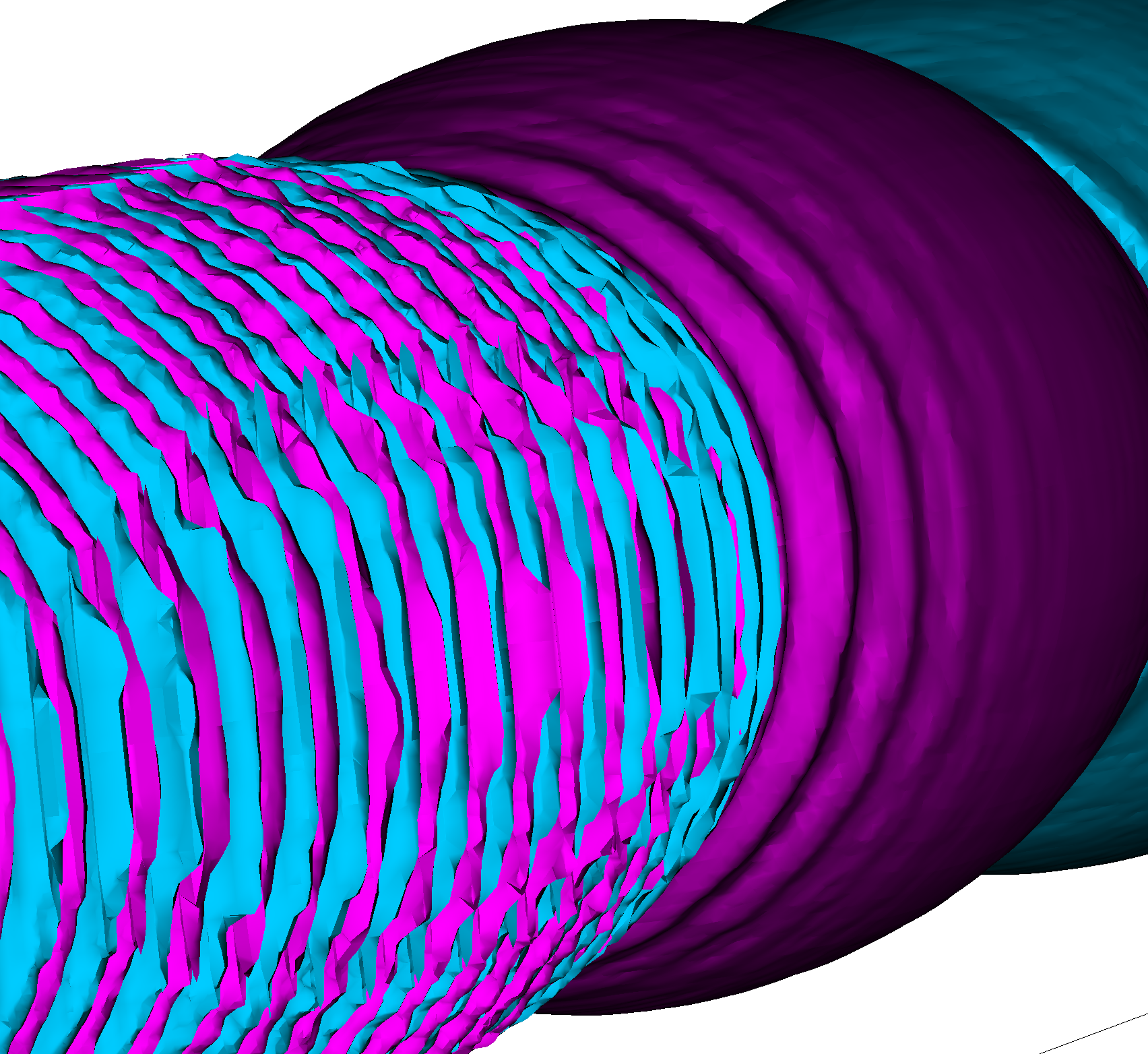}
         \caption{Re-sampling, eb=1E-2, R-SSIM=4.01E-4}
         \label{fig:wpx-dere-2}
     \end{subfigure}
     \begin{subfigure}[t]{0.32\linewidth}
         \centering
         \includegraphics[width=\linewidth]{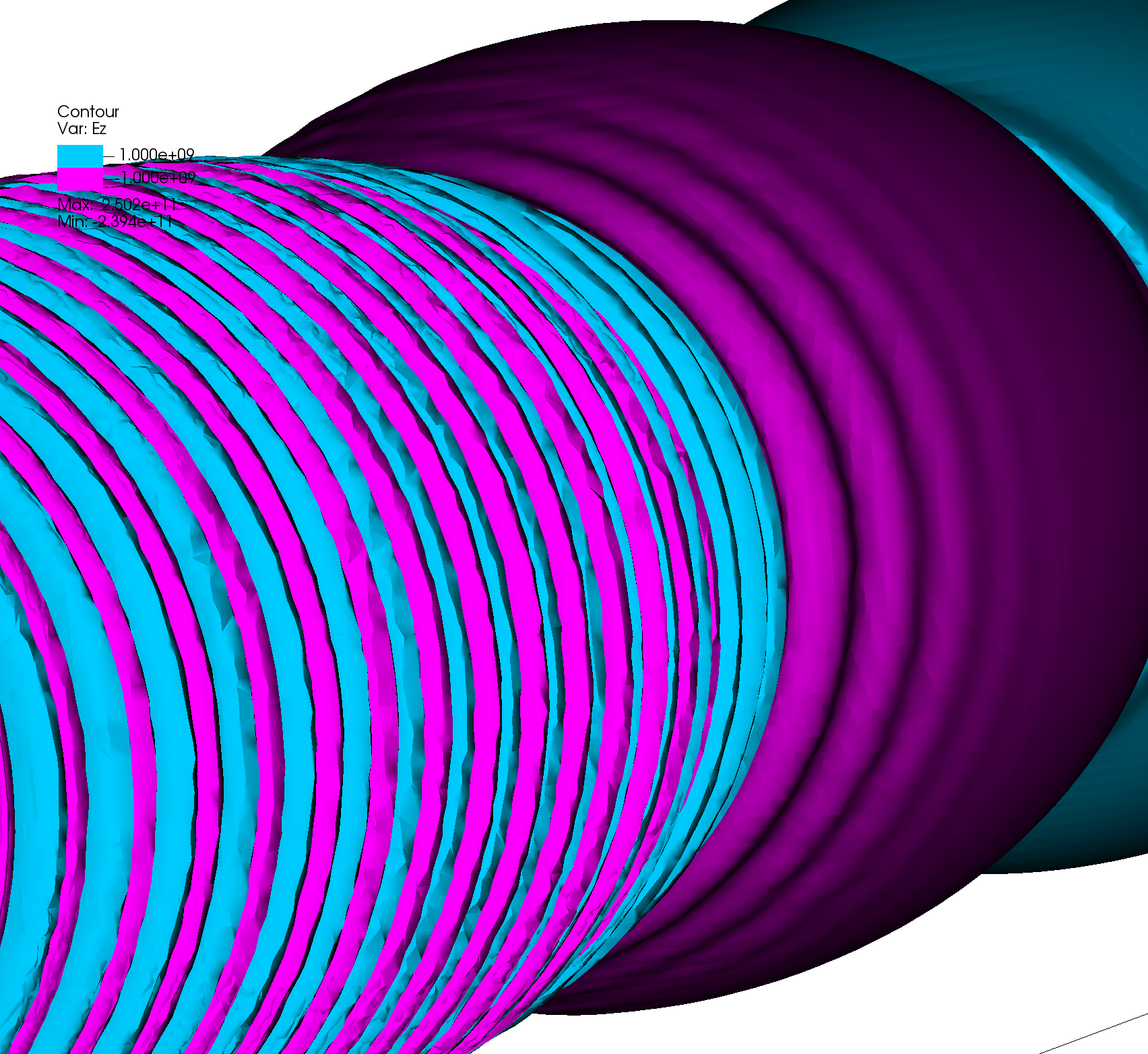}  
         \caption[t]{Dual-cell, eb=1E-4, R-SSIM=2.20E-07}
         \label{fig:wpx-dual-4}
     \end{subfigure} 
     \hspace{0.01\linewidth}
     \begin{subfigure}[t]{0.32\linewidth}
         \centering
         \includegraphics[width=\linewidth]{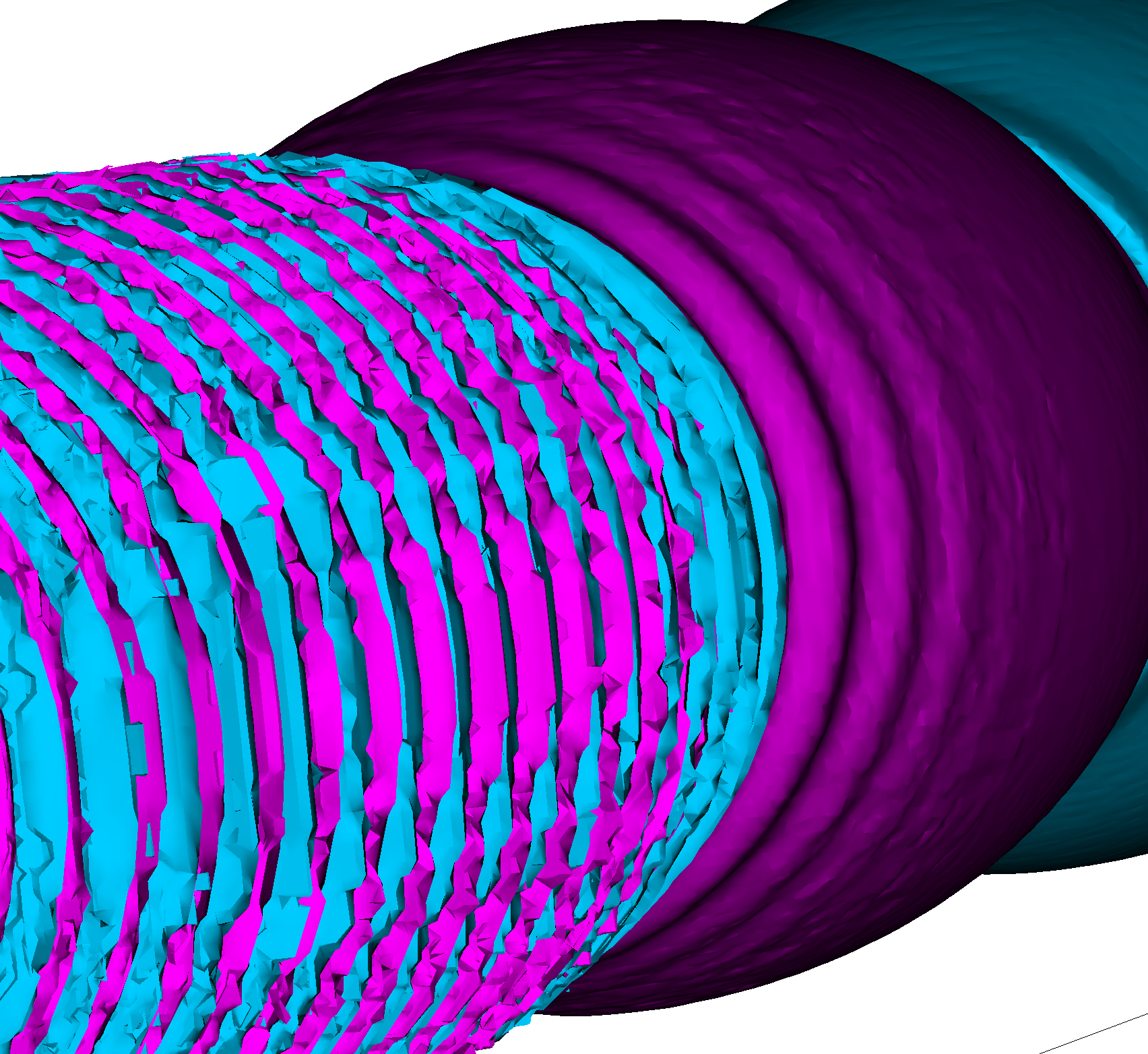}
         \caption{Dual-cell, eb=1E-3, R-SSIM=1.38E-05}
         \label{fig:wpx-dual-3}
     \end{subfigure} 
     \hspace{0.01\linewidth}
     \begin{subfigure}[t]{0.32\linewidth}
         \centering
         \includegraphics[width=\linewidth]{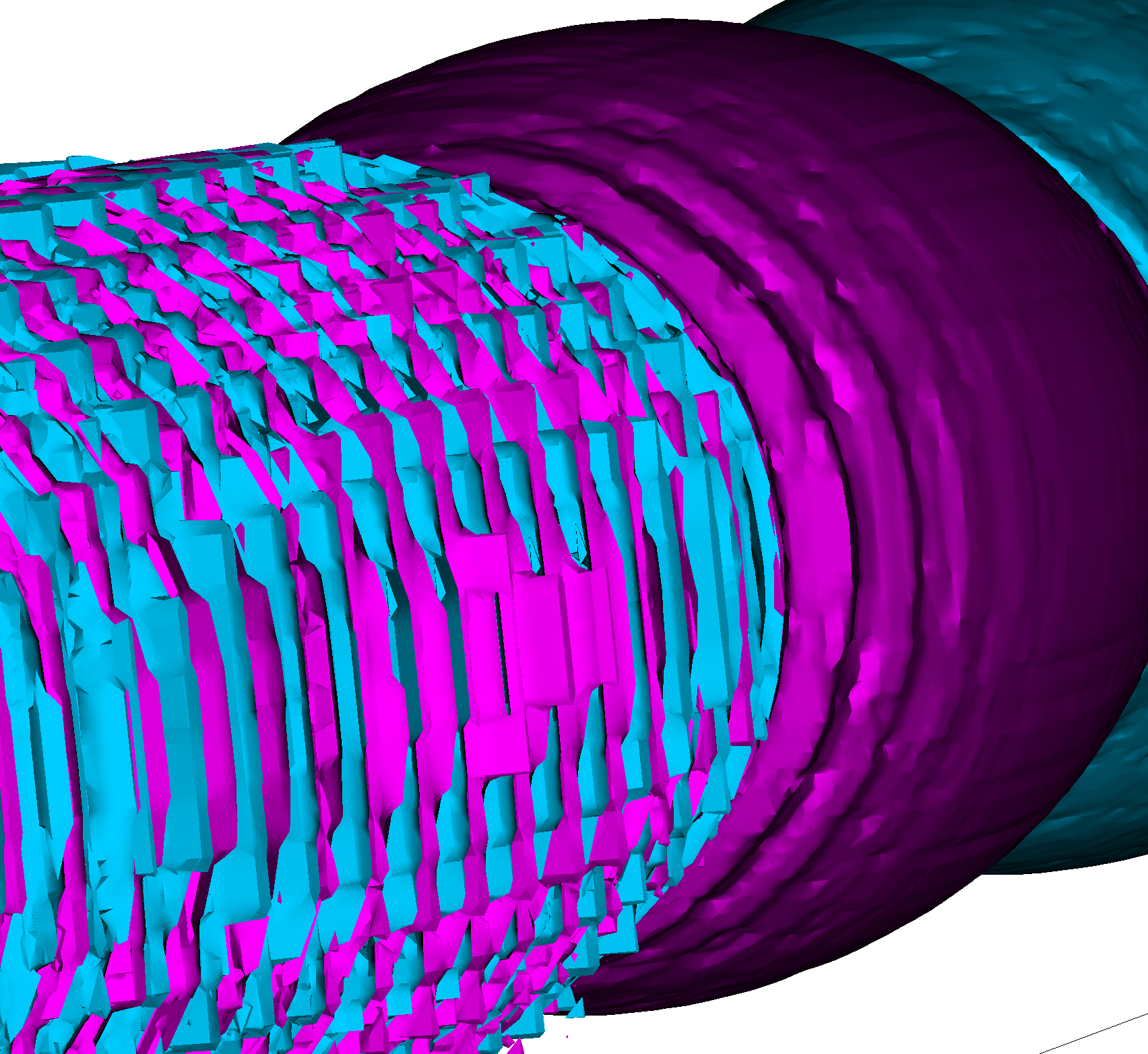}
         \caption{Dual-cell, eb=1E-2, R-SSIM=4.01E-4}
         \label{fig:wpx-dual-2}
     \end{subfigure}
        \caption[t]{
        Iso-surface visualization of compressed AMR data on WarpX using the re-sampling and dual-cell methods (which eliminate cracks between different levels) with different relative error configurations. The compression algorithm employed in these cases is SZ-L/R. See Table~\ref{tab:metric} for detailed quantitative metrics.}
        \label{fig:wpx-sz2}
\end{figure*}

Our evaluation primarily focuses on two AMR applications developed by the AMReX framework~\cite{zhang2019amrex}: Nyx cosmology simulation~\cite{nyx} and the WarpX~\cite{warpx} electromagnetic and electrostatic Particle-In-Cell (PIC) simulation. 
 
Nyx is a cutting-edge cosmology code that employs AMReX and combines compressible hydrodynamic equations on a grid with a particle representation of dark matter. Nyx generates six fields, including baryon density, dark matter density, temperature, and velocities in the $x$, $y$, and $z$ directions. WarpX is a highly parallel and optimized code that utilizes AMReX, runs on GPUs and multi-core CPUs, and features load-balancing capabilities. WarpX can scale up to the world's largest supercomputer and was the recipient of the 2022 ACM Gordon Bell Prize \cite{warpx-gordon}.

We chose these two applications because the data generated by them is very different and thus more representative. Data generated by WarpX is more smooth while the data produced by Nyx appears irregular.
The specifics of our test dataset are presented in Table~\ref{tab:dataset}. In the WarpX application, the grid sizes across levels transition from coarse to fine, featuring dimensions of $128\times128\times1024$ and $256\times256\times2048$, and corresponding data densities of 91.4\% and 8.6\%, respectively. Conversely, for the Nyx application, the grid sizes are $256\times256\times256$ and $512\times512\times512$, associated with data densities of 40.7\% and 59.3\%, respectively.

\subsection{Compression Algorithms}
\label{sec:method-compression}
This work is based on two distinct SZ compression algorithms, namely the SZ-L/R, which employs the Lorenzo and linear regression predictors as delineated in \cite{sz18}, and the SZ-Interp, which utilizes the spline interpolation approach as discussed in \cite{zhao2021optimizing}, as mentioned in Section~\ref{sec:background}. The SZ-L/R algorithm commences by partitioning the entire input data into 6$\times$6$\times$6 blocks, followed by the independent application of either the Lorenzo predictor or high-dimensional linear regression on each individual block. Conversely, the SZ-Interp algorithm conducts interpolation across all three dimensions of the entire dataset.

A key distinction between SZ-L/R and SZ-Interp is that the former is block-based whereas the latter is global. Specifically, the SZ-L/R algorithm divides data into blocks prior to compression, whereas the SZ-Interp algorithm applies global interpolation across the entire dataset. This leads to a notable difference in terms of visualization; the block-based nature of SZ-L/R tends to produce "block-wise" artifacts, as opposed to SZ-Interp. However, by prioritizing local features via block-wise prediction, the SZ-L/R algorithm demonstrates better performance on unsmooth and irregular datasets like Nyx, which will be demonstrated in detail in Section~\ref{sec:finding}.

Additionally, the block-based structure of SZ-L/R confers advantages in terms of cache locality and facilitates random access support. This means that it is possible to partially visualize the data as needed, which helps conserve time and memory footprint since there is no data dependency between individual blocks.


\section{Experiment Results}
\label{sec:finding}
 
\subsection{WarpX}
\begin{figure}[h]
     \centering
     \begin{subfigure}[t]{0.49\linewidth}
         \centering
         \includegraphics[width=\linewidth]{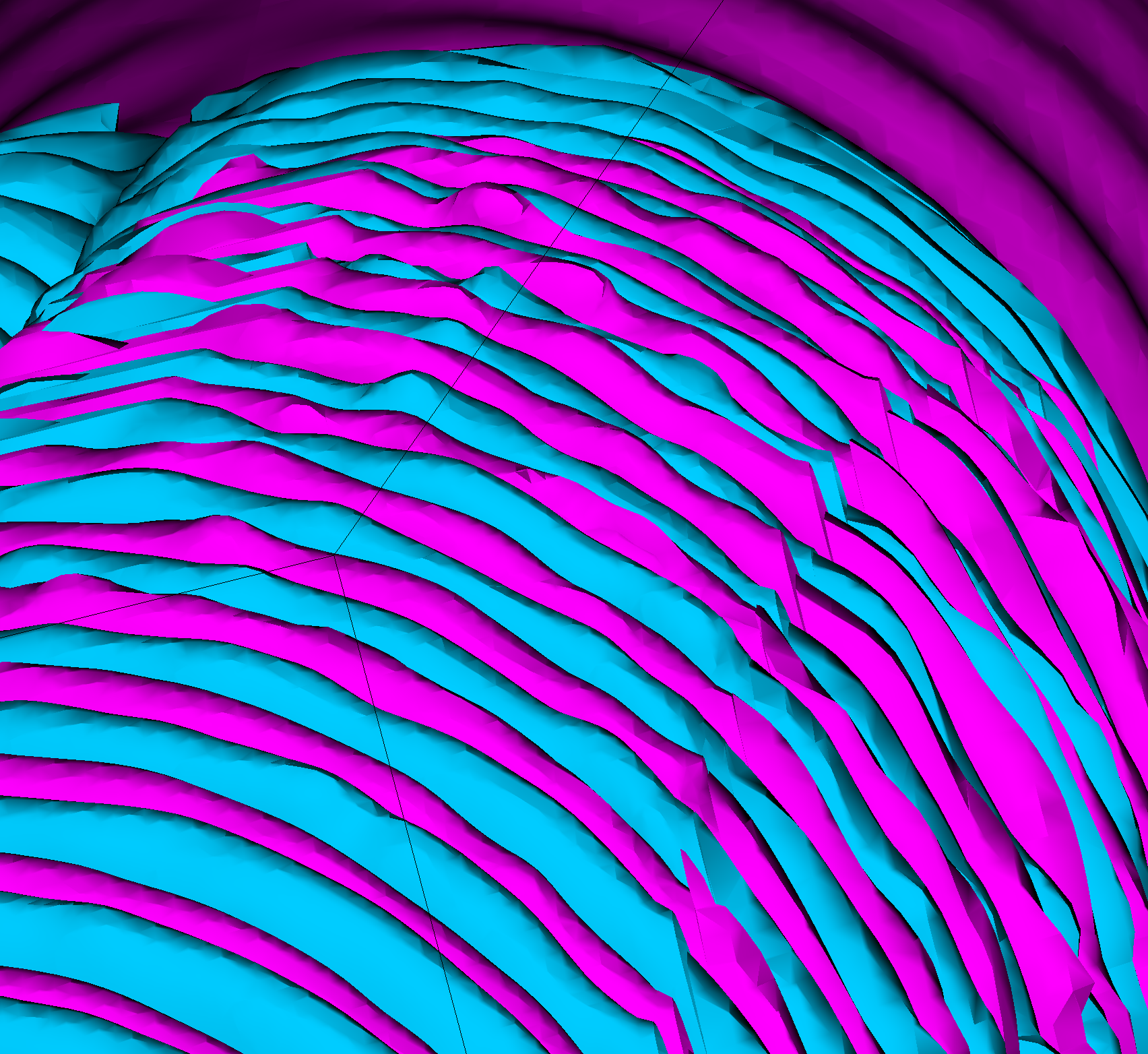}  
         \caption[t]{Re-sampling, eb=1E-3}
         \label{fig:dere-sz3-3}
     \end{subfigure} 
     \begin{subfigure}[t]{0.49\linewidth}
         \centering
         \includegraphics[width=\linewidth]{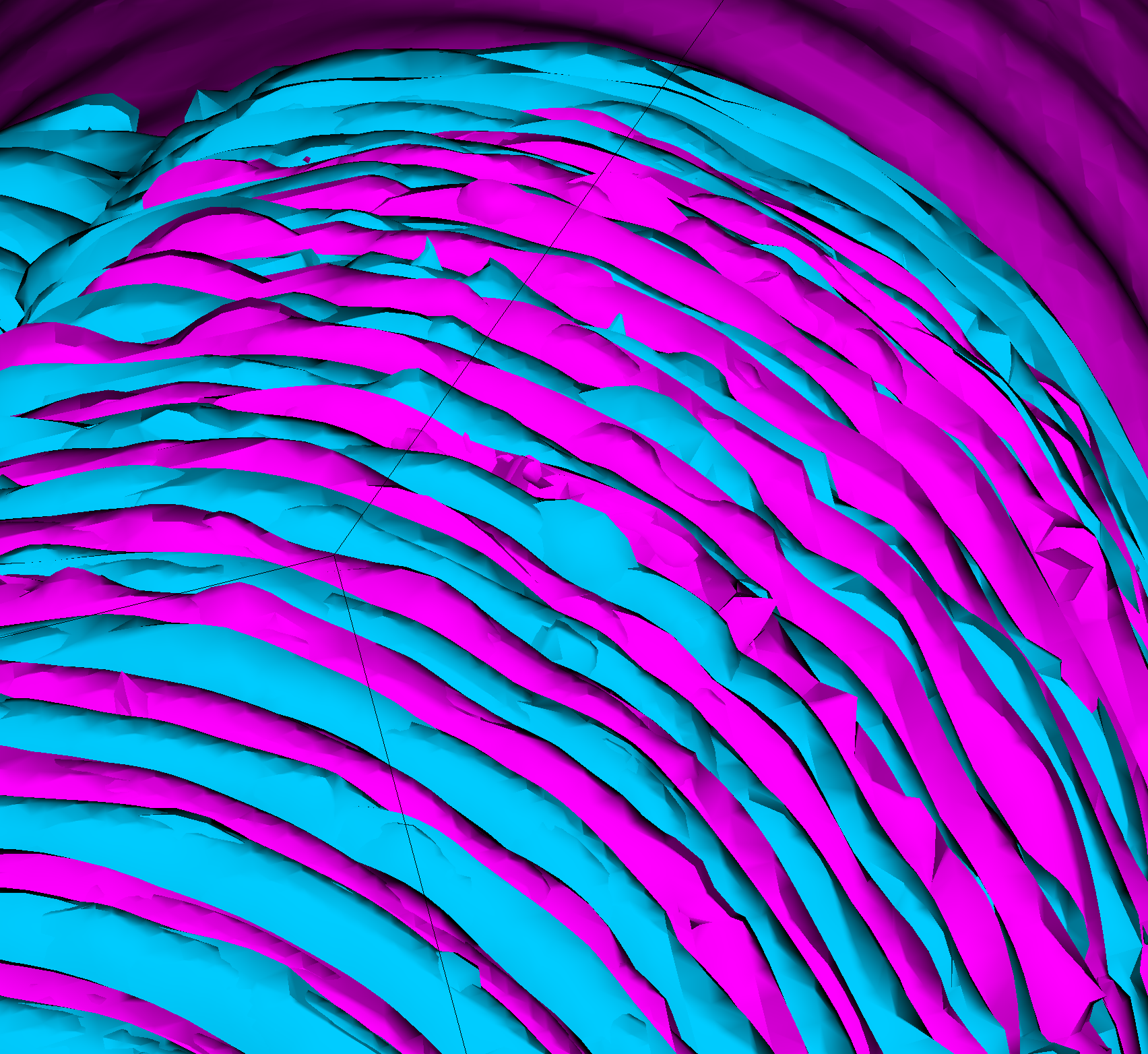}
         \caption{Dual-cell, eb=1E-3, R-SSIM=4.5E-05}
         \label{fig:dual-sz3-3}
     \end{subfigure} 
     \vspace{-2mm}
        \caption[t]{
        The Iso-surface visualization of compressed AMR data is performed using the re-sampling and dual-cell methods, with varying error bounds. The compression algorithm employed in these cases is SZ-Interp. See Table~\ref{tab:metric} for detailed quantitative metrics.}
        \label{fig:wpx-sz3}
\end{figure}

\begin{figure*}[ht]
     \centering
     \begin{subfigure}[t]{0.33\linewidth}
         \centering
         \includegraphics[width=\linewidth]{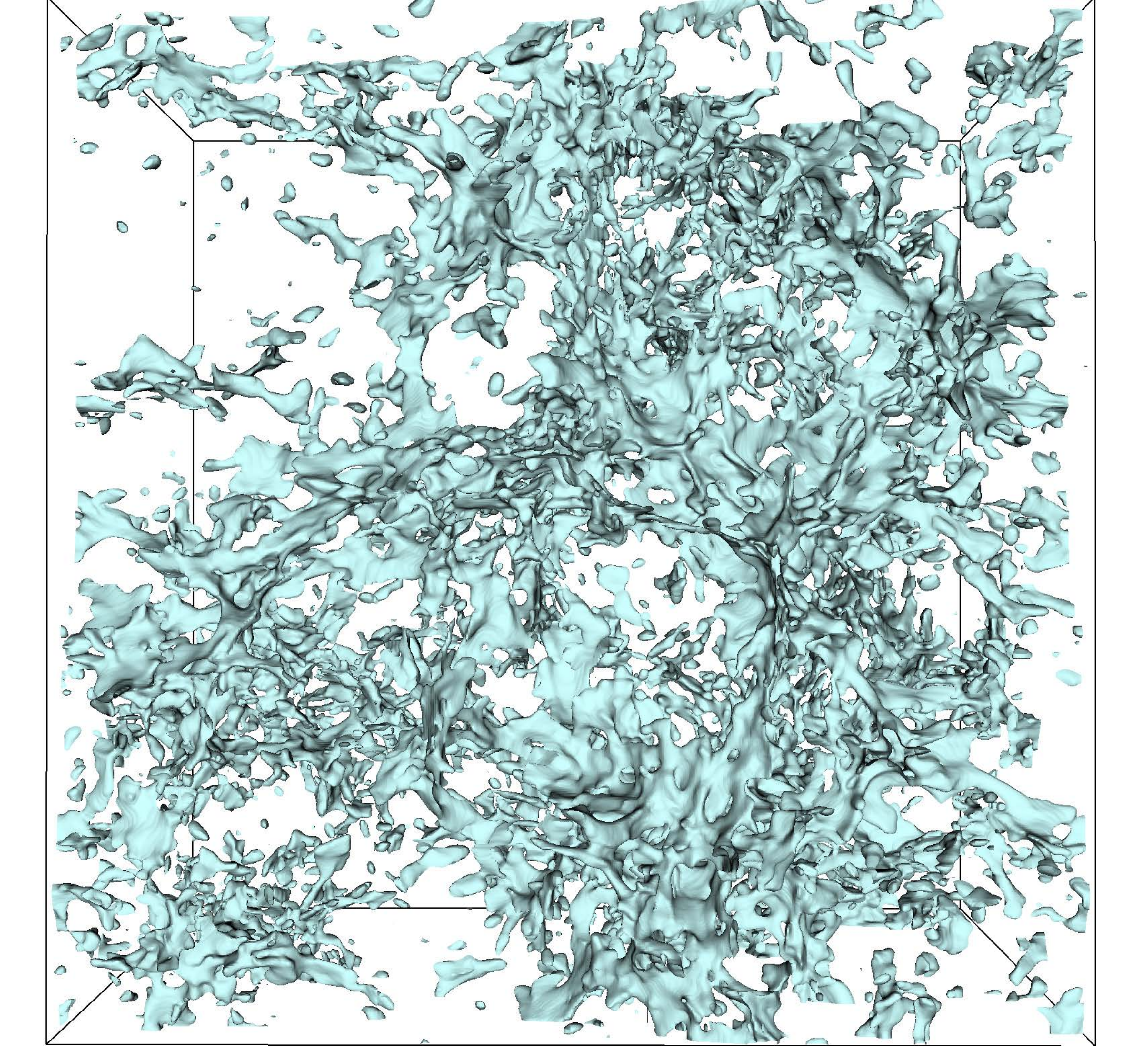}  
         \caption[t]{Re-sampling, Original data}
         \label{fig:nyx-re-ori}
     \end{subfigure} 
     \begin{subfigure}[t]{0.33\linewidth}
         \centering
         \includegraphics[width=\linewidth]{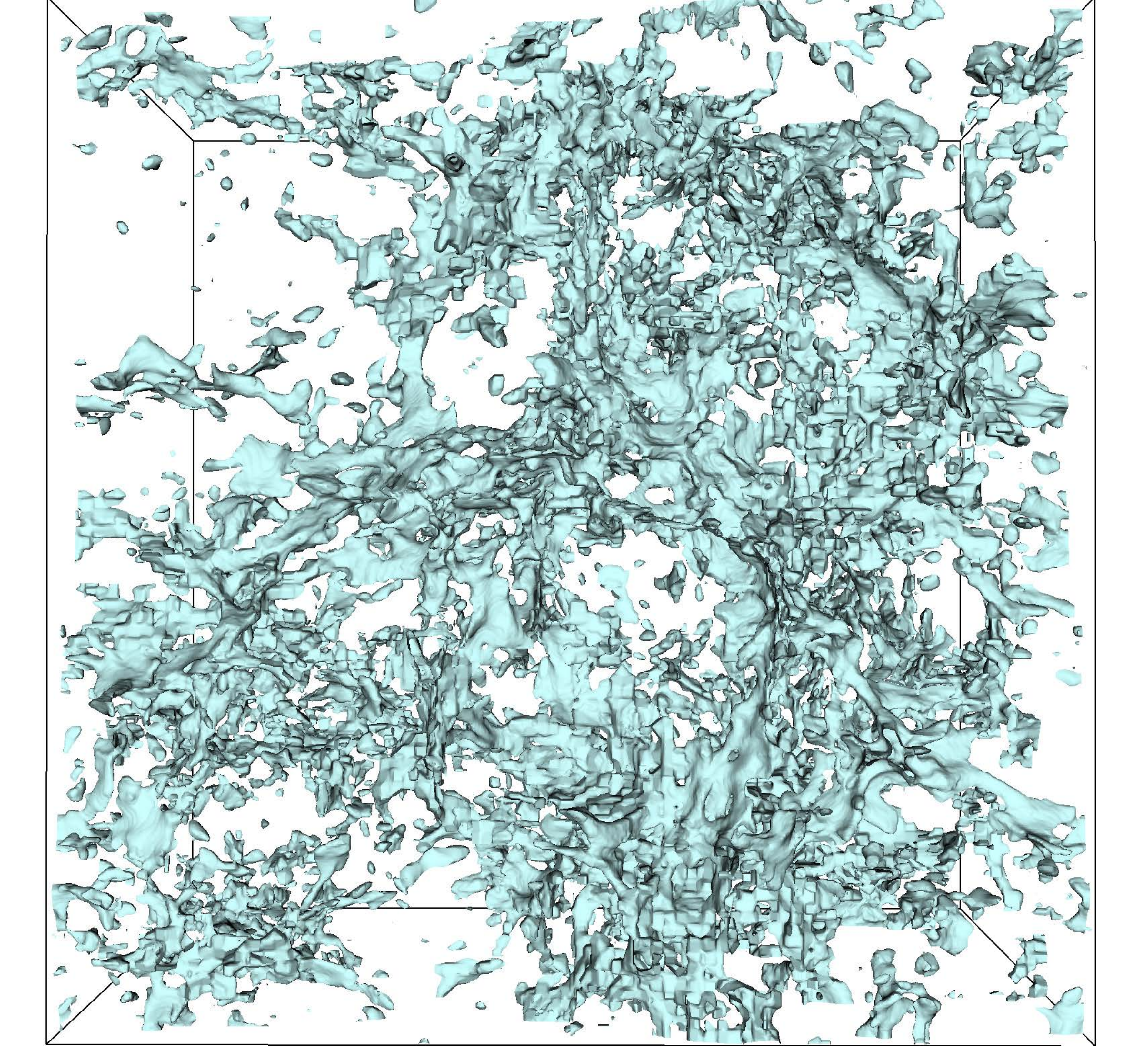}
         \caption{Re-sampling, SZ-L/R, R-SSIM=1.14E-05}
         \label{fig:nyx-re-sz2}
     \end{subfigure} 
     \begin{subfigure}[t]{0.33\linewidth}
         \centering
         \includegraphics[width=\linewidth]{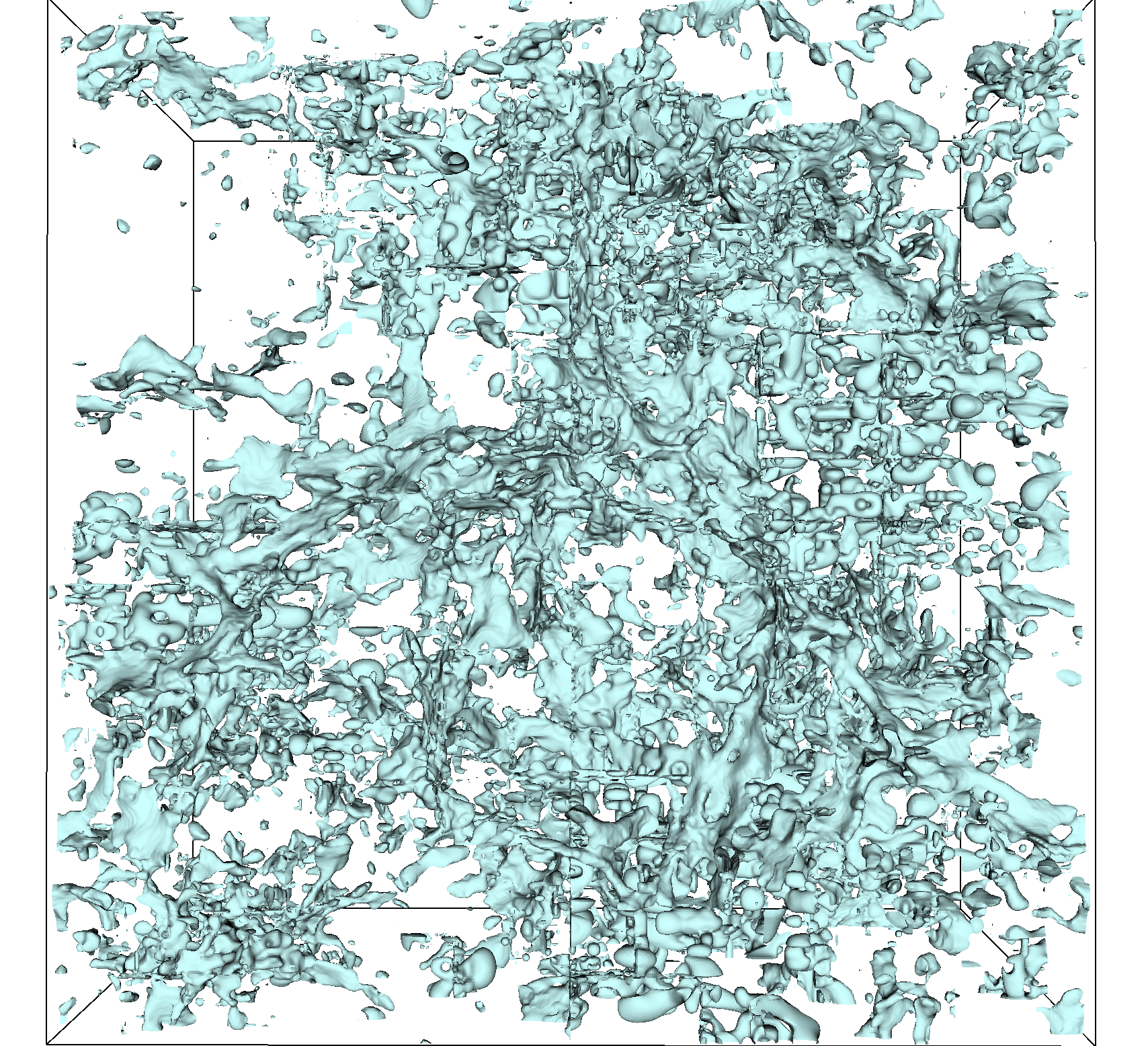}
         \caption{Re-sampling, SZ-Interp, R-SSIM=2.78E-04}
         \label{fig:nyx-re-sz3}
     \end{subfigure}
     \begin{subfigure}[t]{0.33\linewidth}
         \centering
         \includegraphics[width=\linewidth]{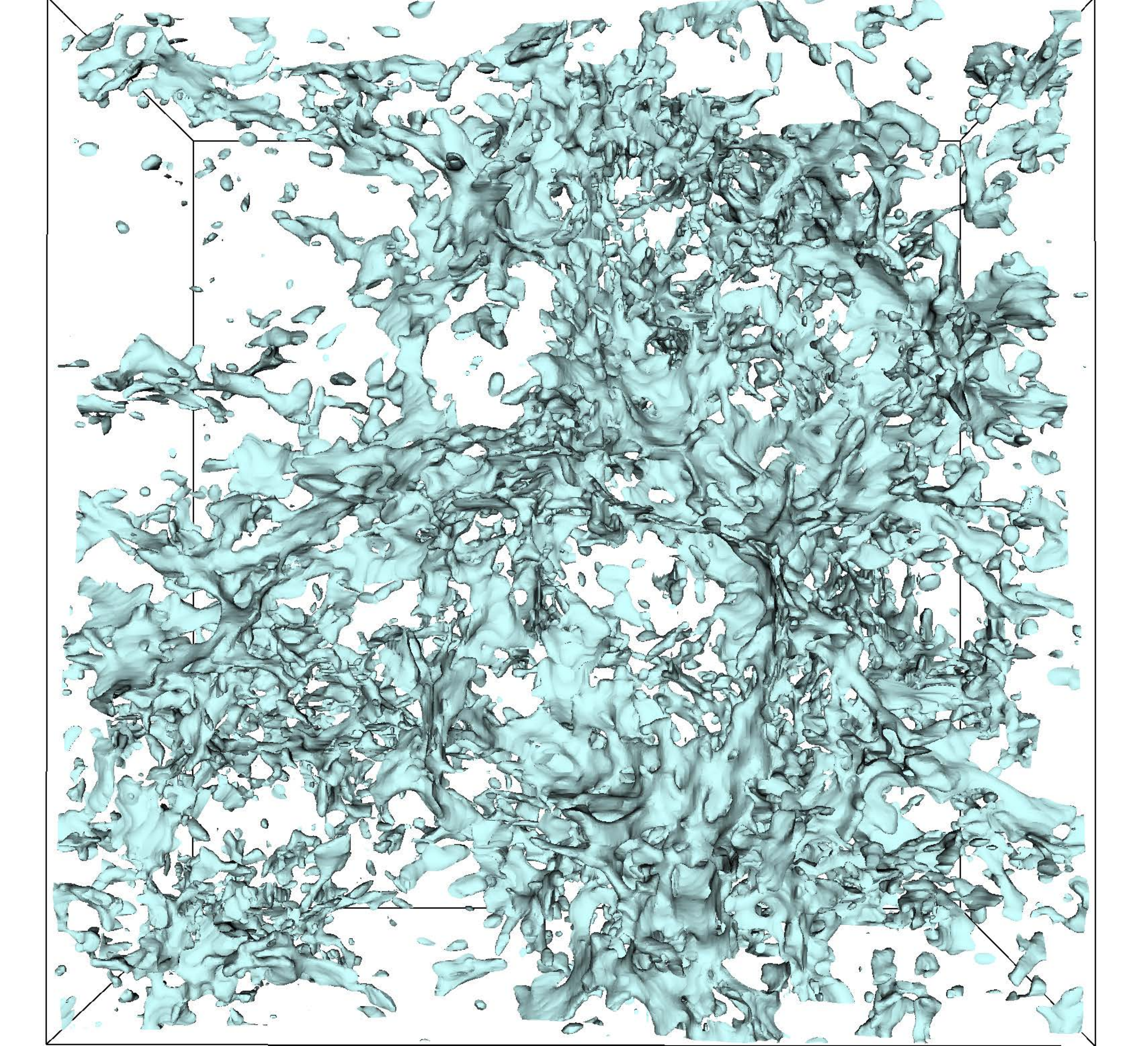}  
         \caption[t]{Dual-cell, Original data}
         \label{fig:nyx-dual-ori}
     \end{subfigure} 
     \begin{subfigure}[t]{0.33\linewidth}
         \centering
         \includegraphics[width=\linewidth]{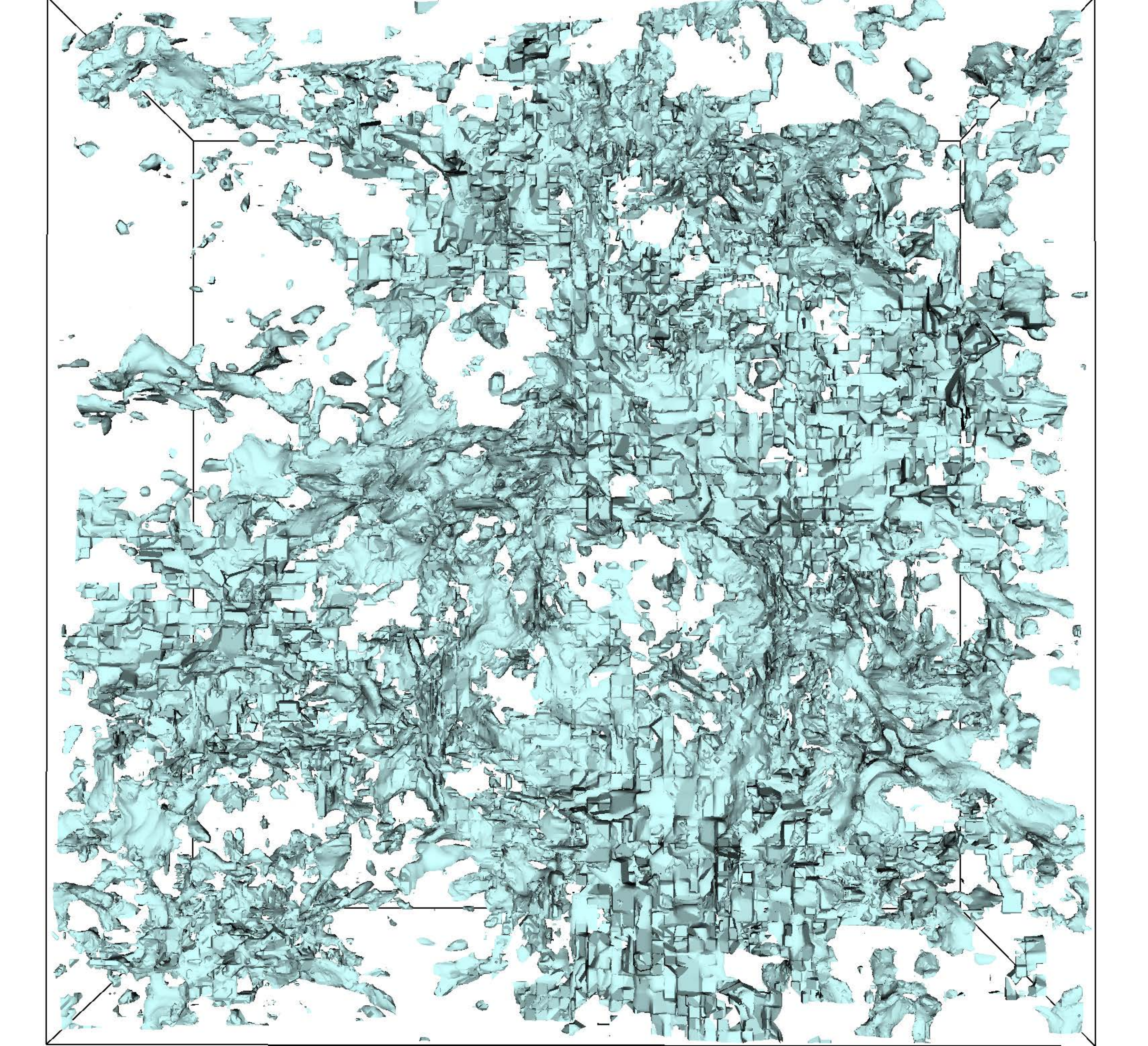}
         \caption{Dual-cell, SZ-L/R, R-SSIM=1.14E-05}
         \label{fig:nyx-dual-sz2}
     \end{subfigure} 
     \begin{subfigure}[t]{0.33\linewidth}
         \centering
         \includegraphics[width=\linewidth]{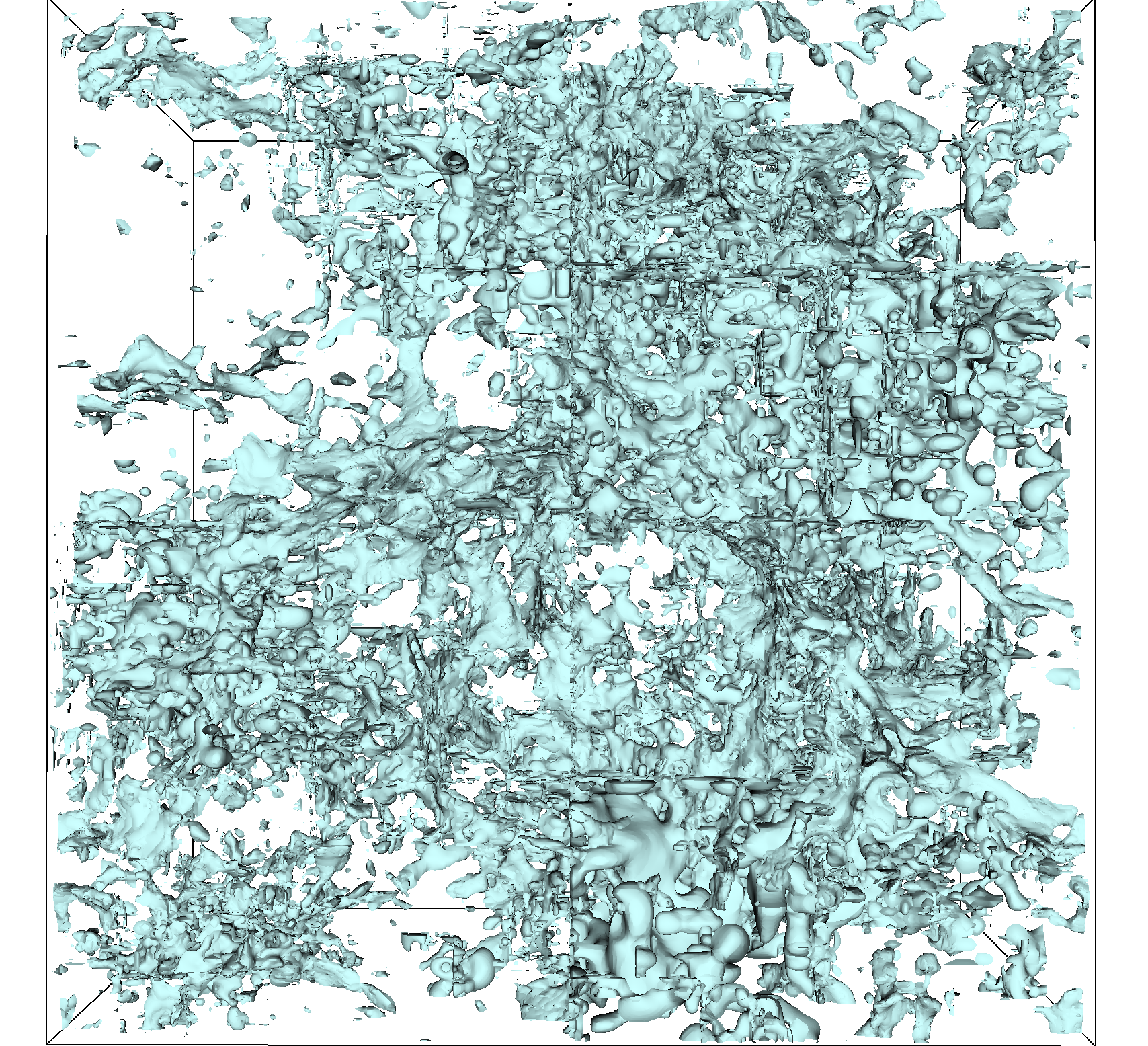}
         \caption{Dual-cell, SZ-Interp, R-SSIM=2.78E-04}
         \label{fig:nyx-dual-sz3}
     \end{subfigure}
     \vspace{-1mm}
        \caption[t]{
        The iso-surface visualization of Nyx AMR data using the re-sampling and dual-cell methods on the original data, as well as the SZ-L/R and SZ-Interp compressed data. The error bounds utilized in these cases are 1E-2. See Table~\ref{tab:metric} for more quantitative metrics.}
        \label{fig:nyx}
\end{figure*}
\begin{table}[h]
\caption{Detailed quantitative result about our experiment.}
\vspace{-2mm}
\resizebox{.99\linewidth}{!}{
\begin{tabular}{|cc|ccc|ccc|}
\hline
\multicolumn{2}{|c|}{Application} &
  \multicolumn{3}{c|}{WarpX} &
  \multicolumn{3}{c|}{Nyx} \\ \hline
\multicolumn{2}{|c|}{Error Bound} &
  \multicolumn{1}{c|}{1E-04} &
  \multicolumn{1}{c|}{1E-03} &
  1E-02 &
  \multicolumn{1}{c|}{1E-04} &
  \multicolumn{1}{c|}{1E-03} &
  1E-02 \\ \hline
\multicolumn{1}{|c|}{\multirow{4}{*}{SZ-L/R}} &
  CR &
  \multicolumn{1}{c|}{23.7} &
  \multicolumn{1}{c|}{31.4} &
  42.3 &
  \multicolumn{1}{c|}{14.6} &
  \multicolumn{1}{c|}{28.6} &
  61.9 \\ \cline{2-8} 
\multicolumn{1}{|c|}{} &
  PSNR &
  \multicolumn{1}{c|}{96.34} &
  \multicolumn{1}{c|}{77.72} &
  60.70 &
  \multicolumn{1}{c|}{102.51} &
  \multicolumn{1}{c|}{90.33} &
  81.09 \\ \cline{2-8} 
\multicolumn{1}{|c|}{} &
  SSIM &
  \multicolumn{1}{c|}{0.999\,999\,8} &
  \multicolumn{1}{l|}{0.999\,986} &
  \multicolumn{1}{l|}{0.999\,60} &
  \multicolumn{1}{c|}{0.999\,999\,9} &
  \multicolumn{1}{l|}{0.999\,998\,8} &
  \multicolumn{1}{l|}{0.999\,989} \\ \cline{2-8} 
\multicolumn{1}{|c|}{} &
  R-SSIM &
  \multicolumn{1}{c|}{2.20E-07} &
  \multicolumn{1}{c|}{1.38E-05} &
  4.01E-4 &
  \multicolumn{1}{c|}{6.87E-08} &
  \multicolumn{1}{c|}{1.20E-06} &
  1.14E-05 \\ \hline
\multicolumn{1}{|c|}{\multirow{4}{*}{SZ-Itp}} &
  CR &
  \multicolumn{1}{c|}{32.4} &
  \multicolumn{1}{c|}{45.1} &
  52.6 &
  \multicolumn{1}{c|}{15.8} &
  \multicolumn{1}{c|}{34.7} &
  77.9 \\ \cline{2-8} 
\multicolumn{1}{|c|}{} &
  PSNR &
  \multicolumn{1}{c|}{96.57} &
  \multicolumn{1}{c|}{78.24} &
  60.38 &
  \multicolumn{1}{c|}{103.11} &
  \multicolumn{1}{c|}{86.63} &
  72.94 \\ \cline{2-8} 
\multicolumn{1}{|c|}{} &
  SSIM &
  \multicolumn{1}{c|}{0.999\,999\,5} &
  \multicolumn{1}{l|}{0.999\,955} &
  \multicolumn{1}{l|}{0.997\,23} &
  \multicolumn{1}{c|}{0.999\,999\,9} &
  \multicolumn{1}{l|}{0.999\,993\,7} &
  \multicolumn{1}{l|}{0.999\,722} \\ \cline{2-8} 
\multicolumn{1}{|c|}{} &
  R-SSIM &
  \multicolumn{1}{c|}{5.19E-07} &
  \multicolumn{1}{c|}{4.50E-05} &
  2.77E-03 &
  \multicolumn{1}{c|}{7.21E-08} &
  \multicolumn{1}{c|}{6.25E-06} &
  2.78E-04 \\ \hline
\end{tabular}
}
\vspace{-1mm}
\label{tab:metric}
\end{table}

We first evaluated the WarpX simulation. As shown in Figure~\ref{fig:wpx-sz2}, the SZ-L/R compression algorithm was applied with different relative error bound ranges from 1E-4 to 1E-2 for both the re-sampling and dual-cell methods. Clearly, the larger the error bound, the worse the visual quality of the data. As mentioned in Section~\ref{sec:back-crack}, the conventional re-sampling method results in a crack between different AMR levels, visible in all the decompressed data in Figures~\ref{fig:wpx-dere-4}, ~\ref{fig:wpx-dere-3}, and ~\ref{fig:wpx-dere-2}.

As also noted in Section~\ref{sec:back-crack}, these cracks can be remedied using the dual-cell method, as demonstrated in Figures~\ref{fig:wpx-dual-4}, \ref{fig:wpx-dual-3}, and \ref{fig:wpx-dual-2}. However, Figures~\ref{fig:wpx-dual-4}, \ref{fig:wpx-dual-3}, and \ref{fig:wpx-dual-2} show that although the dual-cell method can fix the cracks between different AMR levels, it significantly affects the visual quality of the decompressed data compared to the re-sampling as shown in Figures~\ref{fig:wpx-dual-4}, \ref{fig:wpx-dual-3}, and \ref{fig:wpx-dual-2}. Specifically, the dual-cell method amplifies the compression artifact of the data, especially when the error bound is relatively larger, as shown in Figure~\ref{fig:wpx-dual-2}, where the block-wise artifact is greatly emphasized. These block-wise artifacts are due to the block-wise nature of the SZ-L/R compression algorithm, as mentioned in Section~\ref{sec:method-compression}.

The negative impact of the dual-cell method on the decompressed data is even noticeable when the error bound is relatively small. For instance, in Figure~\ref{fig:wpx-dere-4}, using the re-sampling method with a relative error bound of 1E-4, the visual quality of the decompressed data remains (almost) lossless compared to using the re-sampling method on the original data in Figure~\ref{fig:defualt-ori}. However, as shown in Figure~\ref{fig:wpx-dual-4}, when using the dual-cell method with a relative error bound of 1E-4, there is noticeable visual distortion in the decompressed data compared to the original data using the dual-cell method in Figure~\ref{fig:dual-ori}.

For SZ-Interp, it is evident that the dual-cell method could still affect the visual quality compared to re-sampling, as shown in Figure~\ref{fig:wpx-sz3}. There are more bump artifacts when using the dual-cell method in Figure~\ref{fig:dual-sz3-3} compared to re-sampling (Figure~\ref{fig:dere-sz3-3})

\begin{figure}[ht]
     \centering
     \begin{subfigure}[t]{\columnwidth}
         \centering
         \includegraphics[width=0.85\linewidth]{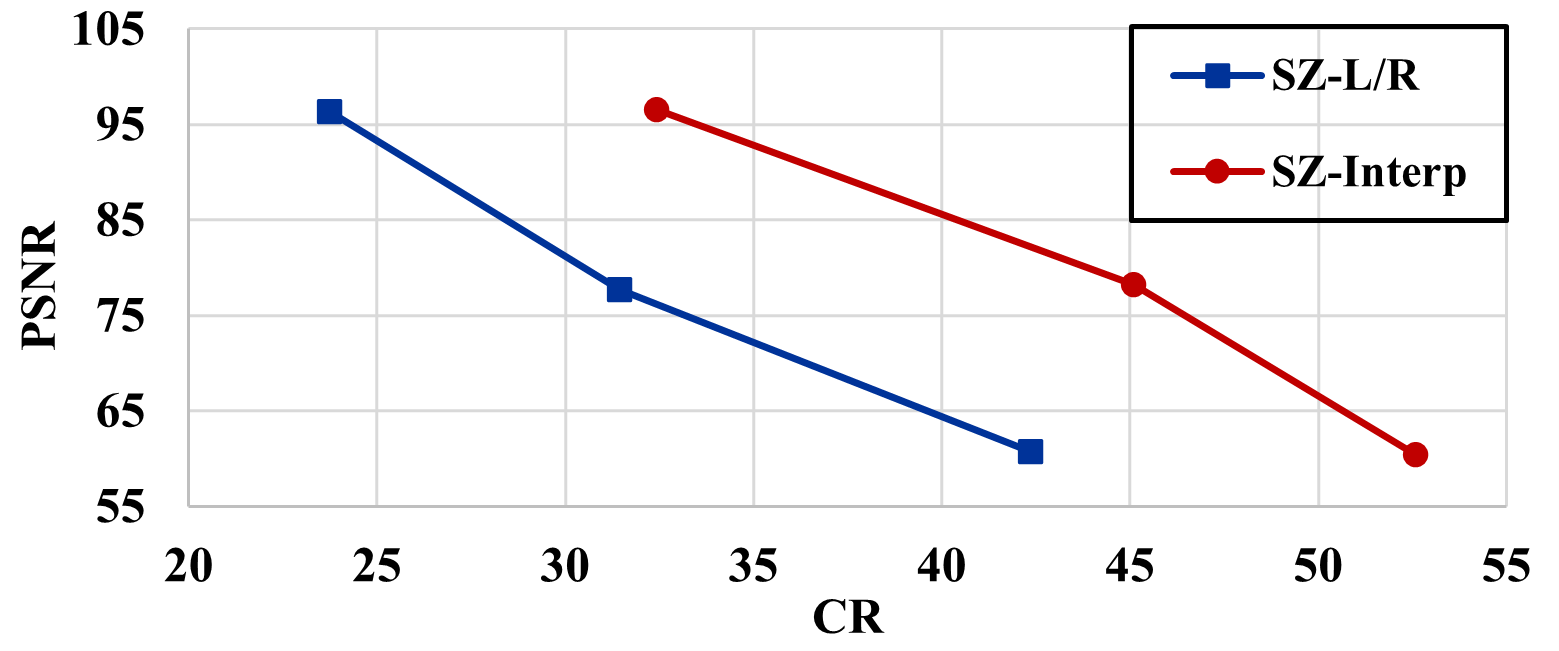}
         \caption{PSNR vs CR}
         \label{fig:wpx-psnr}
     \end{subfigure}
     \begin{subfigure}[t]{\columnwidth}
         \centering
         \includegraphics[width=0.85\linewidth]{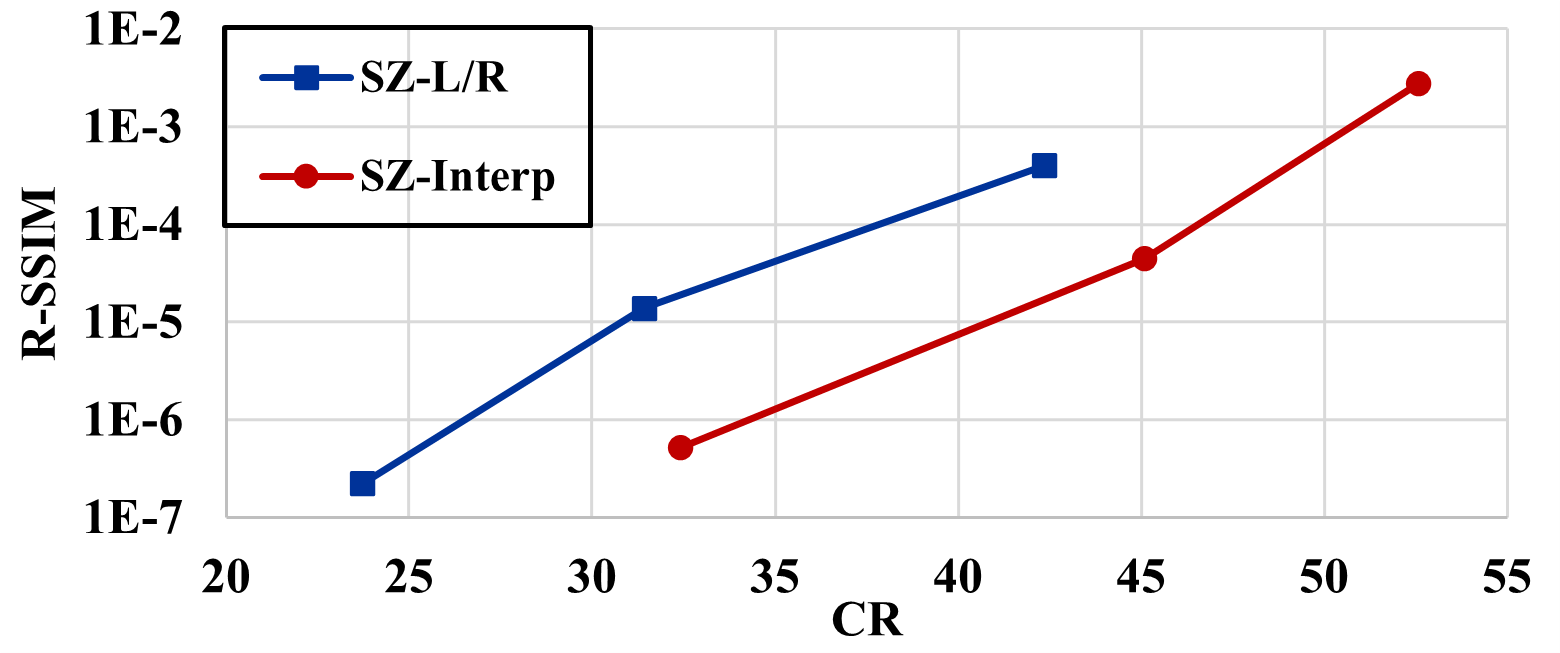}
         \caption{R-SSIM ($1~-~SSIM$) vs CR in log scale}
         \label{fig:wpx-ssim}
     \end{subfigure}
     \vspace{-2mm}
        \caption[t]{Rate-distortion comparison (including both R-SSIM and PSNR) between SZ-L/R and SZ-Interp on WarpX "Ez" field.
        }
        \vspace{-2mm}
        \label{fig:sz2}
\end{figure}

\paragraph{Quantitative metrics} The detailed quantitative metrics are displayed in Table~\ref{tab:metric}. Additionally, as illustrated in Figure~\ref{fig:wpx-psnr} and~\ref{fig:wpx-ssim}, it is evident that SZ-Interp results in better rate-distortion performance in terms of both SSIM and PSNR. This can be primarily attributed to the smooth nature of the data generated by WarpX, which is conducive for the global interpolation predictor of SZ-Interp to perform optimally. 
It is important to note, however, that the SSIM does not provide an intuitive representation in this context because the relative SSIM difference between different error configurations is too small (e.g., 0.999,999,8 for eb=1E-4 vs 0.999,60 for eb=1E-2), despite there being a noticeable difference in the actual visual quality (as can be observed in Figure~\ref{fig:wpx-dual-4} and Figure~\ref{fig:wpx-dual-2}). Therefore, we propose the use of the reverse SSIM (denoted as R-SSIM):

\begin{equation}
R\text{-SSIM} = 1 - SSIM
\end{equation}

In our scenario, the R-SSIM proves to be a more intuitive metric compared to the SSIM, as it more accurately reflects the differences across various cases. As can be seen from Figure~\ref{fig:wpx-sz2} and Table~\ref{tab:metric}, the R-SSIM is intuitive and highly related to the image quality.



\subsection{Nyx}

We then evaluated using the Nyx simulation. As depicted in Figure~\ref{fig:nyx}, we visualized the original data and decompressed data generated by SZ-L/R as well as SZ-Interp for both the re-sampling (Figure~\ref{fig:nyx-re-ori},~\ref{fig:nyx-re-sz2}, and~\ref{fig:nyx-re-sz3}) and dual-cell methods (Figure~\ref{fig:nyx-dual-ori},~\ref{fig:nyx-dual-sz2}, and~\ref{fig:nyx-dual-sz3}). It is still evident that the dual-cell method diminishes the visual quality of the compressed AMR data for both SZ-L/R and SZ-Interp.

Specifically, when comparing with the original data for SZ-L/R, the dual-cell method, as shown in Figure~\ref{fig:nyx-dual-sz2}, results in more severe block-wise artifacts in the lower mid part of the figure, while the block-wise artifact is less obvious using re-sampling, as can be seen in Figure~\ref{fig:nyx-re-sz2}. For SZ-Interp, as depicted in the lower mid part of Figure~\ref{fig:nyx-dual-sz3}, using the dual-cell method leads to more fault geometry compared to re-sampling in Figure~\ref{fig:nyx-re-sz3}.

Another interesting observation is that although SZ-L/R leads to block-wise artifacts, it yields better visualization quality compared to SZ-Interp. As illustrated in Figures~\ref{fig:nyx-re-sz3} and ~\ref{fig:nyx-dual-sz3}, although the decompressed data from SZ-Interp is smooth, it can greatly distort the original data structure in areas with complex geometry (e.g., the lower mid part).
On the other hand, the block-wise nature of SZ-L/R can help it better capture the complex local patterns/information of the data, leading to better compression and visual quality on the complex and irregular Nyx dataset.

\begin{figure}[ht]
     \centering
     \begin{subfigure}[t]{\columnwidth}
         \centering
         \includegraphics[width=0.85\linewidth]{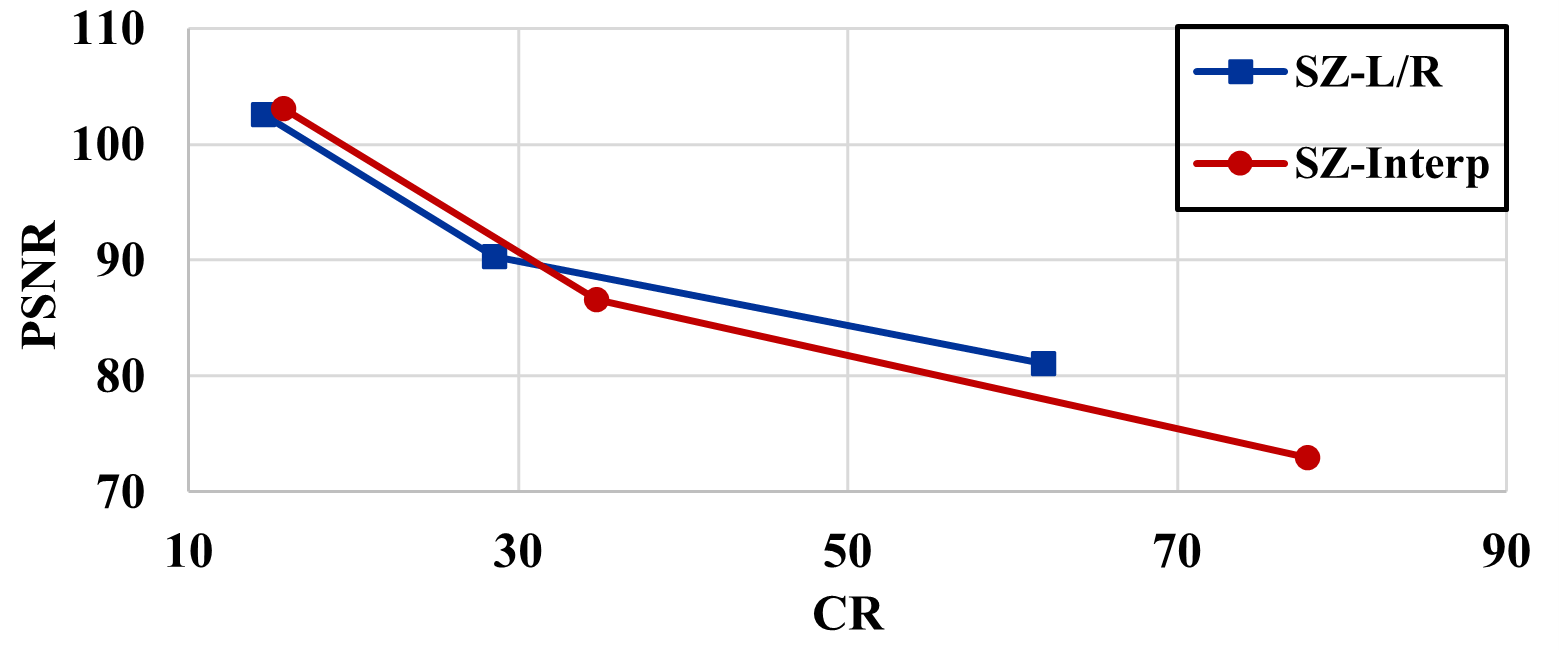}
         \caption{PSNR vs CR}
         \label{fig:nyx-psnr}
     \end{subfigure}
     \begin{subfigure}[t]{\columnwidth}
         \centering
         \includegraphics[width=0.87\linewidth]{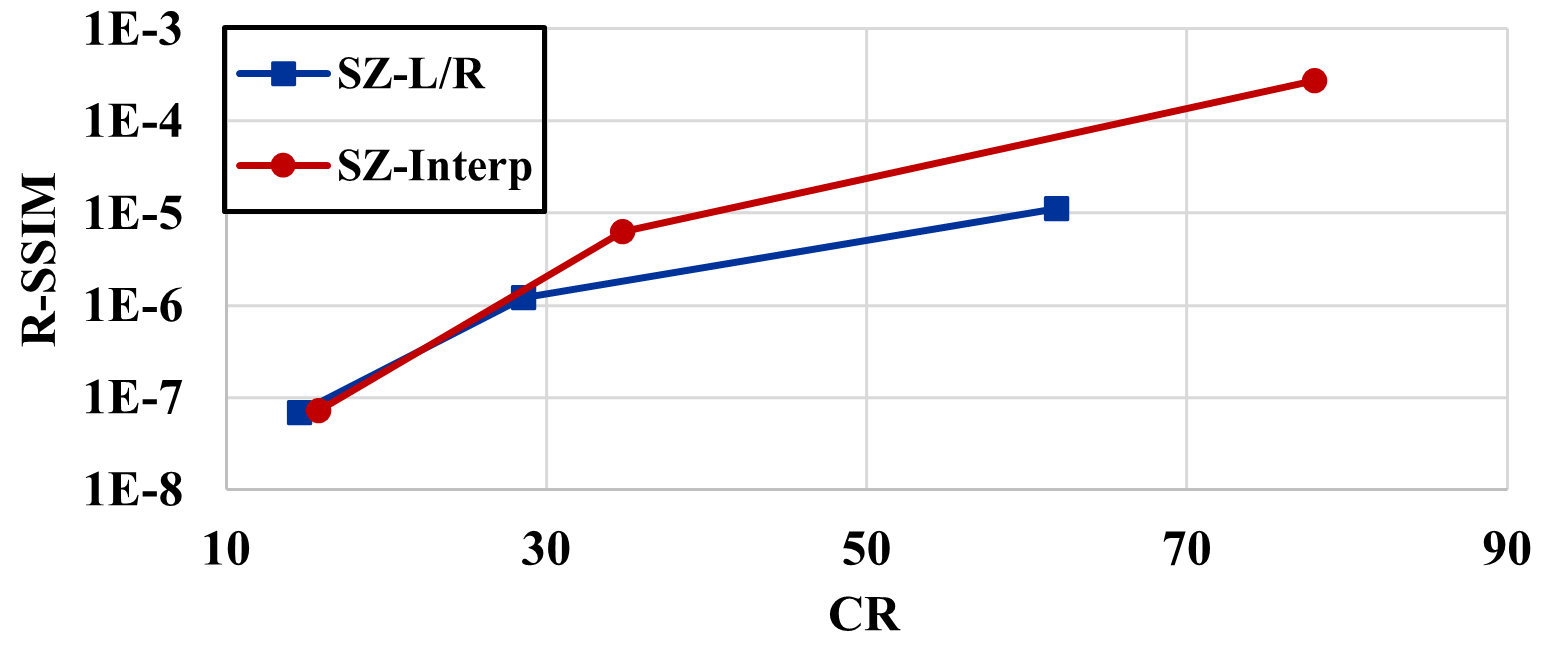}
         \caption{R-SSIM ($1~-~SSIM$) vs CR in log scale}
         \label{fig:nyx-ssim}
     \end{subfigure}
     \vspace{-2mm}
        \caption[t]{Rate-distortion comparison (including both R-SSIM and PSNR) between SZ-L/R and SZ-Interp on Nyx "Density" field.
        }
        \vspace{-2mm}
        \label{fig:sz2}
\end{figure}

\paragraph{Quantitative metrics} The detailed quantitative metrics are shown in Table~\ref{tab:metric}. As shown in Figure~\ref{fig:nyx-psnr} and~\ref{fig:nyx-ssim}, for the Nyx dataset, SZ-Interp does not exhibit a superior performance in rate-distortion compared to SZ-L/R. In fact, SZ-L/R even outperforms in terms of R-SSIM (a lower R-SSIM means a higher SSIM). This observation aligns with our previous finding that the block-based predictor of SZ-L/R is more adept at capturing local patterns within the Nyx data.

\subsection{Discussion}

Intuitively, compared to the dual-cell method, the higher iso-surface visualization quality of compressed data with the re-sampling seems to be attributed to the increased resolution of the dataset through interpolation during re-sampling, as discussed in Section~\ref{sec:re-mc}. This interpolation results in a smoother iso-surface. However, for a larger dataset, the resolution advantage of interpolation is relatively minimal. For a dataset with dimensions of $n^3$, the advantage can be quantified as $(n+1)/n$. Given that our AMR dataset has a maximum fine level size of $512^3$, the interpolation during re-sampling only yields a $513/512 = 0.2\%$ higher resolution. This minor advantage explains the visual similarities between the original uncompressed data visualized using dual-cell and re-sampling methods, as depicted in Figures~\ref{fig:defualt-ori} and~\ref{fig:dual-ori} for WarpX and Figure~\ref{fig:nyx-re-ori} and~\ref{fig:nyx-dual-ori} for Nyx, except for the cracks between AMR levels. However, this marginal increase in resolution does not account for the significant negative impact of the dual-cell method on compressed data.

\begin{figure}[t]
     \centering
      \includegraphics[width=0.85\linewidth]{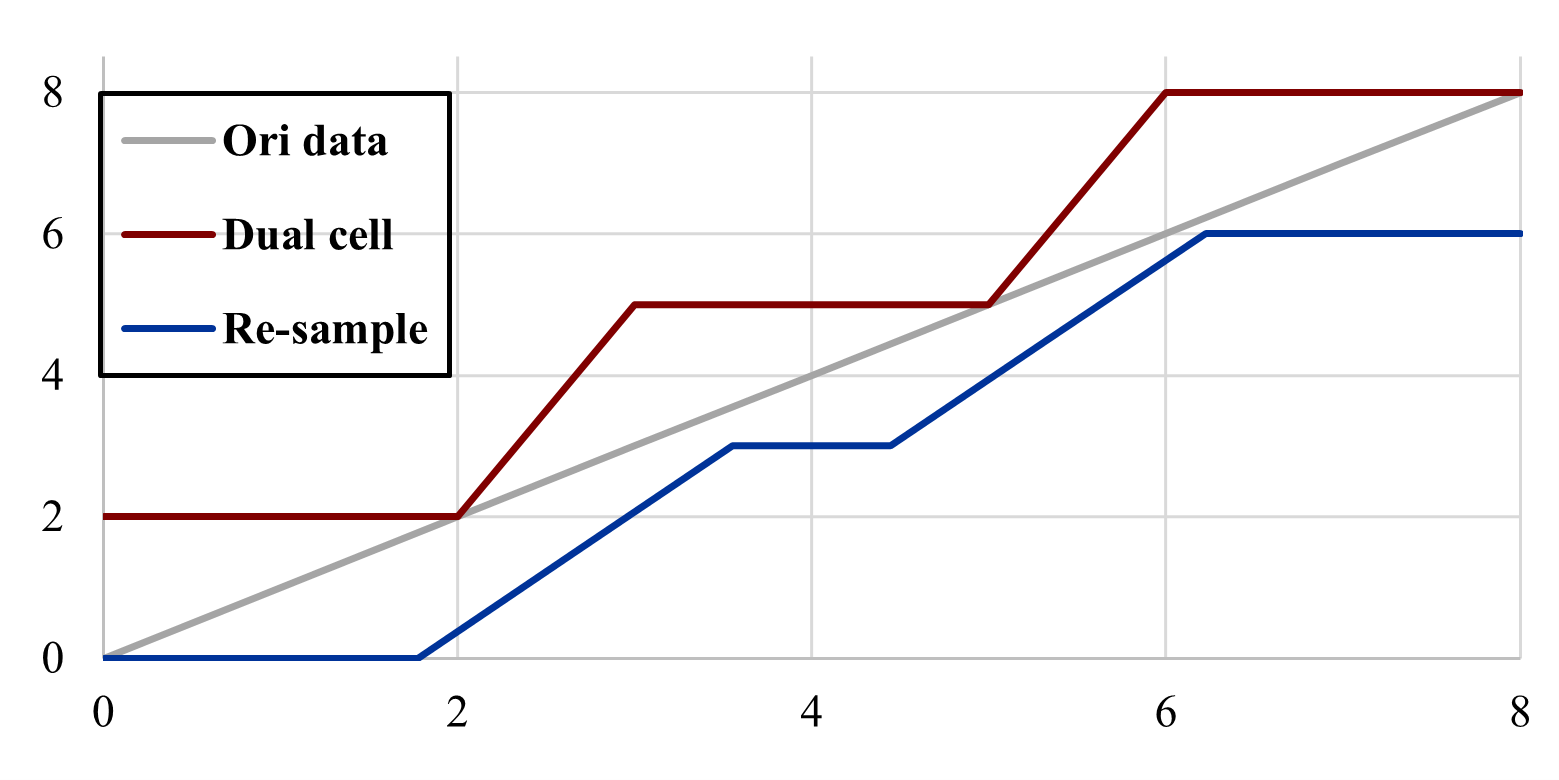}
      \vspace{-2mm}
        \caption[t]{1D visualization example of the original data, decompressed data using dual cell, and decompressed data using re-sampling with interpolation. The decompressed data are shifted by 1 and -1 for more clear demonstration.}
        \label{fig:why}
\end{figure}

Our understanding is that the significant visual contrast between dual-cell and re-sampling methods on decompressed data arises because interpolation can reduce compression artifacts, especially the blocking artifacts of SZ-L/R. Figure~\ref{fig:why} illustrates this with a 1D example. Suppose the original data is represented by the gray curve ``012345678'' and the decompressed data from SZ-L/R is ``111//444//777'', with ``111'', ``444'', and ``777'' indicating block-wise artifacts from the compression (shown as the red curve). Re-sampling and interpolation transform the decompressed data into ``111//\textbf{2.5}//44//\textbf{5.5}//777'' where ``2.5'' and ``5.5'' are interpolated values that mitigate the block artifacts (represented by the blue curve).

\section{Conclusion and Future Work}
\label{sec:conclusion}
In conclusion, the article examines the impact of data compression on Adaptive Mesh Refinement (AMR) data visualization, a subject that has not been thoroughly explored in previous research. While error-bounded lossy compression is identified as a crucial solution to manage the vast amounts of data generated by scientific simulations, its impact on AMR data, which is particularly challenging to visualize due to its hierarchical and multi-resolution nature, is not well understood. 
The article aims to fill this knowledge gap by investigating how data compression not only affects but also introduces new challenges to AMR data visualization, thus contributing to a more comprehensive understanding of AMR data visualization with compression.
\begin{acks}
This work has been authored by employees of Triad National Security, LLC, which operates Los Alamos National Laboratory under Contract No. 89233218CNA000001 with the U.S. Department of Energy (DOE) and the National Nuclear Security Administration (NNSA). 
This research was supported by the Exascale Computing Project (ECP), Project Number: 17-SC-20-SC, a collaborative effort of the DOE SC and NNSA.
This work was also supported by NSF awards 2303064, 2247080, 2311876, and 2312673.
This research used resources of the National Energy Research Scientific Computing Center, a DOE SC User Facility located at Lawrence Berkeley National Laboratory, operated under Contract No. DE-AC02-05CH11231. We would like to thank Dr. Zarija Lukić and Dr. Jean Sexton from the NYX team at Lawrence Berkeley National Laboratory for granting us access to cosmology datasets. 
This research used the open-source particle-in-cell code WarpX \url{https://github.com/ECP-WarpX/WarpX}, primarily funded by the US DOE Exascale Computing Project. Primary WarpX contributors are with LBNL, LLNL, CEA-LIDYL, SLAC, DESY, CERN, and TAE Technologies. We acknowledge all WarpX contributors.
\end{acks}

\bibliographystyle{ACM-Reference-Format}
\bibliography{refs}

\end{document}